\documentclass[twocolumn,
showpacs,
showkeys,
preprintnumbers,
nofootinbib,
superscriptaddress,
amsmath,
amssymb,
floatfix,
secnumarabic,
aps,
pra,
a4paper,
notitlepage,
final,
]{revtex4-1}%

\usepackage[colorlinks=true,urlcolor=blue]{hyperref}
\usepackage{graphicx}
\usepackage{epsfig}

\newcommand{\beq}{\begin{equation}}
\newcommand{\eeq}{\end{equation}}
\newcommand{\beqar}{\begin{eqnarray}}
\newcommand{\eeqar}{\end{eqnarray}}
\newcommand{\bal}{\begin{aligned}}
\newcommand{\eal}{\end{aligned}}

\def\ham{\hbox{$\cal H$}}
\def\dalam{\hbox
{\vrule\vbox{\hrule\hbox to 1ex{ \hfill}\kern 1 ex\hrule}\vrule}}

\def\1/2{\hbox{$ {1 \over 2}$ }}
\def\tr{\hbox{Tr}}

\def\h{\hbar}
\def\i/h{{i \over \h}}

\def\ch{\cosh}
\def\sh{\sinh}

\def\inf{\infty}
\def\pd{\partial} 
\def\v{\vec}

\def\cvE{\hbox{ $ \vec {\cal { E }}$}} \def\cvH{\hbox{$ \vec {\cal { H}}$}}

\def\a{\alpha}  
\def\b{\beta}  
 
\def\g{\gamma} \def\G{\Gamma} 
\def\d{\delta} \def\D{\Delta} 
\def\l{\lambda}  
\def\e{\epsilon} \def\E{\hbox{$\cal E $}}

\def\r{\rho} \def\vr{\varrho}

\def\p{\psi} \def\P{\Psi}
\def\bp{\bar \psi}

\def\m{\mu}
\def\n{\nu}
\def\t{\tau}
  \def\vk{\varkappa}

\def\z{\zeta}

\def\tt{\theta} \def\vt{\vartheta}

\def\<{\langle}
\def\>{\rangle}

\def\({\left(}
\def\[{\left[}
\def\){\right)}
\def\]{\right]}

\usepackage{subfigure}

\usepackage[notcite,notref,color]{showkeys}  

\usepackage{tikz}
\usetikzlibrary{decorations.pathreplacing}
\usetikzlibrary{decorations.pathmorphing}    
\usepackage{multirow}
\usepackage{dcolumn}		
\newcolumntype{.}{D{.}{.}{-1}}
\newcolumntype{i}[1]{D{.}{.}{#1}}

\newcommand{\myfrac}[2]{{\ifmmode{}^{#1}\!/_{\!#2}\else${}^{#1}\!/_{\!#2}$\fi}}

\begin{document}
\sloppy

\title{Spontaneous emission in the supercritical QED: what is wrong and what is possible}

\author{P.Grashin and K.Sveshnikov}
\email{k.sveshnikov@gmail.com}
\email{p.grashin@physics.msu.ru}
\affiliation{Department of Physics and
N.N.Bogoliubov Lab of Theoretical and Mathematical Physics, Moscow State
University, 119991, Leninsky Gory, Moscow, Russia}


\date{\today}


\begin{abstract}

   Spontaneous positron emission, caused by the supercritical Coulomb source with charge $Z$ and size $R$, is explored in essentially non-perturbative approach  with emphasis on the  VP-energy $\E_{VP}$\footnote{Throughout the paper the abbreviation "VP"\, stands for "vacuum polarization"\,.}, considered as a function of the Coulomb source parameters $Z$ and $R$.   The specific contribution to $\E_{VP}$, which appears in the supercritical case due to  direct Coulomb interaction  between VP-charge densities $\vr_{VP}(\v r)$, is outlined and studied in detail. It is shown that  after first levels diving into the lower continuum  this contribution, being properly renormalized,  turns out to be negative in contrast to the classical electrostatics  and so could play an important  role in spontaneous  emission, especially for $Z$ just above the first $Z_{cr,1}$.   The additional problems of spontaneous  emission, caused by the lepton  number conservation, are also discussed.
\end{abstract}


\maketitle

\section{Introduction}

 So far, the behavior of QED-vacuum exposed to a supercritical EM-source is subject to an active research~\cite{Rafelski2016,Kuleshov2015a,*Kuleshov2015b,*Godunov2017,Davydov2017,Sveshnikov2017,
Popov2018,*Novak2018,*Maltsev2018,Roenko2018,Maltsev2019,*Maltsev2020}.
Of the main interest is the assumption that  in such external fields there should take place  a deep vacuum state reconstruction, caused by discrete levels diving into the lower continuum and accompanied by such nontrivial effects as spontaneous positron emission combined with vacuum  shells formation (see e.g., Refs.~\cite{Greiner1985a,Plunien1986,Greiner2012,Ruffini2010,Rafelski2016} and citations therein). In 3+1 QED,  such effects are expected for extended Coulomb sources of nucleus size with charges $Z>Z_{cr,1} \simeq 170$, which are large enough for direct  observation and probably could be created in low energy heavy ion collisions  at new heavy ion facilities like  FAIR (Darmstadt), NICA  (Dubna), HIAF (Lanzhou)~\cite{FAIR2009,Ter2015,MA2017169}.

The problem of spontaneous positron emission in the supercritical QED has a long story, starting from the pioneering works of Frankfurt (W.Greiner, B.Mueller, J.Rafelsky, et al.) and Moscow (Ya.Zeldovich, S.Gerstein, V.Popov, et al.) groups (see, e.g., Refs.~\cite{Rafelski2016,Greiner1985a,Zeldovich1972} and citations therein) in the early seventies. However, in the subsequent investigations of supercritical heavy ion collisions at  GSI (Darmstadt, Germany), rechecked later by Argonne Lab. (USA), no evidence of the diving phenomenon was found ~\cite{Mueller1994}. The next generation of accelerator facilities is expected to drive these investigations to a new level~\cite{FAIR2009,Ter2015,MA2017169}. The novel experimental study requires an updated  theoretical analysis.

 In the present paper the non-perturbative VP-effects, caused by the quasi-static supercritical Coulomb source with $Z>Z_{cr,1}$ and size $R$, are explored in terms of VP-energy $\E_{VP}$ and vacuum shells formation. $\E_{VP}$ plays an essential role in  the region of super-criticality, especially for spontaneous  emission, since the latter should be provided solely by the VP-effects  without any other channels of energy transfer. In particular, it is indeed the decline of $\E_{VP}$, which provides the spontaneous positrons with corresponding energy for emission.
  Being considered as a function of $Z$, VP-energy reveals with growing $Z$ a pronounced decline into the negative range, accompanied with negative jumps, exactly equal to the electron rest mass, which occur each time when the  discrete level dives into the lower continuum~\cite{Grashin2022a}.  At the same time, being considered as a function of $R$ with fixed $Z$,  $\E_{VP}(R)$ simulates in a quite reasonable way the non-perturbative VP-effects in  slow heavy ion collisions. In view of recent attempts in this field of interest
~\cite{Rafelski2016,Kuleshov2015a,*Kuleshov2015b,*Godunov2017,Popov2018,*Novak2018,*Maltsev2018,
Roenko2018,Maltsev2019,*Maltsev2020,FAIR2009,Ter2015,MA2017169}, these circumstances require for a special study.

Up to now, the main attention was paid to the VP-energy of the polarized Dirac sea~\cite{Greiner1985a,Plunien1986,Greiner2012,Grashin2022a}
\beq
\label{1.1}
\E_{D,VP}=\frac{1}{2}\(\sum\limits_{\e_{n}<\e_{F}} \e_n  -   \sum\limits_{\e_{n}\geqslant \e_{F}} \e_n \) \ ,
\eeq
where $\e_F=-m_e\,c^2$ is the Fermi level, which in such problems with strong Coulomb fields is chosen at the lower threshold, while $\e_{n}$  are the eigenvalues  of the corresponding Dirac-Coulomb (DC) problem. Actually, $ \E_{D,VP} $ is nothing else but the Casimir energy for the electron-positron system in the external Coulomb field~\cite{Plunien1986}.

However, it is not the total VP-energy of the system. Namely,  in the Coulomb gauge, within  which we are dealing, there appears an additional contribution to the total VP-energy, reproducing the direct Coulomb interaction between vacuum charge densities $\vr_{VP}(\v r)$,\footnote{We are really surprised  that this term has not yet been considered in the context of the supercritical QED-effects.}
\beq
\label{1.2}
\E_{C,VP}=\frac{1}{2} \int \! d{\v r}\,d{\v r'}\, {\vr_{VP}(\v r) \ \vr_{VP}(\v r') \over |\v r - \v r'|} \ .
\eeq
The origin of this term is the corresponding operator-valued expression, which appears within the canonical QED-quantization in the Coulomb gauge (see, e.g., Ref.~\cite{Bjorken1965}, Chapter 15). So the ultimate expression for the total VP-energy under question reads
\beq
\label{1.3}
\E_{VP}= \E_{D,VP} + \E_{C,VP} \ .
\eeq

The purpose of this work is to deal indeed with the expression (\ref{1.3}) in order to obtain  the correct  behavior of VP-energy in the supercritical region for $Z>Z_{cr,1}$ with special attention to the contribution of the Coulomb term $\E_{C,VP}$. Two main reasons should be outlined here. First, in the supercritical region the main contribution to VP-charge density $\vr_{VP}(\v r)$ is produced via vacuum shells formation, which change the total VP-charge $Q_{VP}$. Without this contribution $\vr_{VP}(\v r)$ remains a small perturbative effect, which cannot significantly alter the whole picture.  Second, the contribution of vacuum shells turns out to be quite different from what could be expected from general grounds. Vacuum shells appear as empty vacancies after level diving into the lower continuum and become negatively charged only after the resonance decay, accompanied with positron emission~\cite{Greiner1985a,Plunien1986,Greiner2012}. Without  positron emission they remain empty and so do not contribute to $\vr_{VP}(\v r)$. At the same time, after they charge by means of positron emission,  $\E_{C,VP}^{ren}$ turns out to be negative in contrast to the classical electrostatics. This effect is quite similar to the perturbative Uehling one, where the spatial distribution of induced by the point Coulomb source VP-density is opposite to the classical picture~\cite{Greiner1985a,Plunien1986,Greiner2012}. And although $\E_{C,VP}^{ren}$ turns out to be much smaller than the main term (\ref{1.1}),  the negative sign of correctly renormalized  $\E_{C,VP}^{ren}$  plays an important  role in  spontaneous emission, especially for $Z$ just above the first $Z_{cr,1}$, where indeed this negative contribution to VP-energy provides an additional amount of kinetic energy for positrons to be emitted with non-vanishing probability.  Unfortunately, even with account for this specific effect of $\E_{C,VP}^{ren}$, the realistic estimates for start-up of positron emission turn out to be not less than $Z^\ast \simeq 250$ (see below)\footnote{At this stage, we intentionally do not discuss the problem of the lepton number.    Positron emission implies that  the   corresponding amount of positive lepton number  should be left as  a density, concentrated in vacuum shells. Otherwise, either the lepton number conservation  in such processes must be broken, or the positron emission  prohibited. But there are no indications that the lepton number can exist in the form of a spatial density. So the lepton number can create an absolute ban on spontaneous emission.}. It should be specially noted that without renormalization the contribution of $\E_{C,VP}$  to the total VP-energy still contains negative jumps at each levels diving point, but afterwards continuous to grow and so yields additional problems for spontaneous emission.

Without any serious loss of generality, such a study can be performed within the  Dirac-Coulomb (DC) problem  with external spherically-symmetric Coulomb potential, created by a uniformly charged sphere
\beq
\label{1.5a}
V(r)=- Q\,\( {1 \over R}\, \tt(R-r)+ {1 \over r}\, \tt(r-R) \)  \ ,
\eeq
or  charged ball
\begin{multline}\label{1.6}
V(r)=- Q\,\( {3\, R^2 - r^2 \over 2\,R^3 }\, \tt(R-r) \ +  \right. \\ \left.  + \ {1 \over r}\, \tt(r-R) \)  \ .
\end{multline}
Here and henceforth
\beq
\label{1.5b}
Q=Z \a \ ,
\eeq
while the radius $R$ of the source varies in the range from
\beq
\label{1.8}
R_{min}(Z) \simeq 1.2\, (2.5\, Z)^{1/3} \ \hbox{fm} \ ,
\eeq
which roughly imitates  the size of a super-heavy nucleus with charge $Z$, up to certain $R_{max}$ of order of one electron Compton length, where the VP-effects are already small and show up as  $O(1/R)$-corrections.

It  should be specially noted that the parameter $Q$ plays actually the role of the effective coupling constant for the VP-effects under question. The size of the source  and its shape are also the additional input parameters, but their role in VP-effects is quite different from $Q$ and in some important questions, in particular, in the renormalization procedure, this difference must be clearly tracked. Furthermore, the difference between the charged sphere and the ball, which seems more preferable as a model of a super-heavy nucleus or heavy-ion cluster, in the VP-effects of  Coulomb super-criticality  is  small. It  shows up mainly in the ratio $ \E_{VP,ball}/\E_{VP,sphere} \simeq 6/5$ for the same $Z$ and $R$ with the only condition that $R$ must be close to $R_{min}(Z)$~\cite{Grashin2022a}. At the same time,   the charged sphere model allows for an almost completely analytical study of the problem, which has clear advantages in many positions. The ball model doesn't share such options, since  explicit solution of the DC problem in this case is absent and so one has to use from beginning the numerical methods or special approximations~\cite{Grashin2022a}.

As in basic works on this topic ~\cite{Wichmann1956,Gyulassy1975, McLerran1975a,*McLerran1975b,*McLerran1975c,Greiner1985a,Plunien1986,Greiner2012,Ruffini2010, Rafelski2016},  radiative corrections from virtual photons are neglected. Henceforth, if it is not stipulated separately, relativistic units  $\hbar=m_e=c=1$ and the standard representation of  Dirac matrices are used. Concrete calculations, illustrating the general picture, are performed for $\a=1/137.036$ by means of Computer Algebra Systems (such as Maple 21) to facilitate  the analytic calculations  and GNU Octave code for boosting the numerical work.

The paper is arranged as follows.  In Sections II and III a consistent study of VP-density and the vacuum shells formation in the supercritical case is presented. In Section IV we consider the evaluation of VP-energy in an essentially  non-perturbative approach with emphasis on renormalization and on the most effective methods of calculation. Section V is devoted to presentation of results in the case of spherically-symmetric  Coulomb source (uniformly charged sphere and ball), the most important of which are shown in Figs.\,\ref{VP(Z=184-200,R)}-\ref{VP(Z=300,R)}. With account of these results, in Section VI the question of what is wrong and what is possible in spontaneous emission is explored. Section VII is devoted to concluding remarks and general discussion.

\section{General problem statement}

The most efficient non-perturbative evaluation of the  VP-charge density $\vr_{VP}(\vec{r})$ is based on the  Wichmann and Kroll (WK) approach ~\cite{Wichmann1956,Gyulassy1975,Mohr1998}. The starting point of the latter is the vacuum value
\beq \label{3.1}
\vr_{VP}(\vec{r})=-\frac{|e|}{2}\(\sum\limits_{\e_{n}<\e_{F}} \p_{n}(\vec{r})^{\dagger}\p_{n}(\vec{r}) -  \sum\limits_{\e_{n}\geqslant \e_{F}} \p_{n}(\vec{r})^{\dagger}\p_{n}(\vec{r}) \) \ .
\eeq
 In (\ref{3.1}) $\e_F=-1$ is the Fermi level, which in such problems with strong Coulomb fields is chosen at the lower threshold, while $\e_{n}$ and $\p_n(\vec{r})$ are the eigenvalues and properly normalized set of  eigenfunctions of the corresponding DC problem. The expression (\ref{3.1}) for the  VP-charge density is a direct consequence of the well-known Schwinger prescription for the  fermionic current in terms of the fermion fields commutators
\beq
\label{3.1a}
 j_{\mu}(\v r, t)=-{|e| \over 2}\, \[ \bp(\v r, t)\,, \g_{\mu} \p(\v r, t)\] \ .
 \eeq

An important point here is that  in the case under question  $\vr_{VP}(\v r)$ turns out to be not an average, but an eigenvalue of the fermionic charge density operator $j_0 (\v r, t)$ acting on the vacuum state. The last statement follows immediately from the expansion of the fermionic fields in terms of creation-annihilation operators in the Furry picture
 \beq
\label{3.1b}
\p(\v r, t)=\sum\limits_{\e_{n}\geqslant \e_{F}}  b_{n}\, \p_{n}(\vec{r})\,\mathrm{e}^{-i\e_{n}t}  + \sum\limits_{\e_{n} < \e_{F}}  d_{n}^{\dagger}\, \p_{n}(\vec{r})\,\mathrm{e}^{-i\e_{n}t}  \ ,
 \eeq
where
\beq
\label{3.1c}
 \{ b_n\,, b_{n'}^{\dagger} \} = \d_{n n'} \ , \quad \{ d_n\,, d_{n'}^{\dagger} \}=\d_{n n'} \ ,
 \eeq
while all the other anticommutators vanish, and the complete  set   $\{\p_{n}(\v r)\}$ is chosen in accordance with definitions given above. Then the charge density operator can be equivalently represented as the sum of its normal-ordered form and the VP-density introduced in (\ref{3.1})
\beq
\label{3.1d}
 j_{0}(\v r, t)=: j_{0}(\v r, t): + \vr_{VP}(\v r)  \ ,
 \eeq
and since by definition the vacuum state $|vac\>$ is subject of relations
\beq
\label{3.1e}
  b_n\, |vac\> = d_n\,|vac\> = 0 \ ,
 \eeq
one obtains
\beq
\label{3.1f}
 j_{0}(\v r, t)\, |vac\> =\vr_{VP}(\v r)\, |vac\>  \ .
 \eeq
Therefore $\vr_{VP}(\v r)$ turns out to be a true c-number component of the charge density. This result should be contrasted with, e.g.,  the classical EM-wavepacket, which appears as an average of the EM-field operator over the coherent state with indefinite number of photons, created by the classical external EM-source. The latter is indeed an average with non-vanishing dispersions, both in electric and  magnetic components of the wavepacket,  which are caused by  fluctuations of the photon number in each oscillator mode.

For quadratic in fermion fields theories like QED and other gauge field models the same statement holds also for the vacuum energy. In QED it happens because the Schwinger prescription for the current (\ref{3.1a}) dictates the following form of the Dirac Hamiltonian in the external EM-field (for details see, e.g., Ref.~\cite{Plunien1986})
\begin{multline}
\label{3.1g}
\ham_D (\v r, t) = \\ = {1\over 4i}\,\Big\{ \[ \p(\v r, t)^{\dagger}\,, \v\a\,  \v \nabla \p(\v r, t)\] + \[ \p(\v r, t)\,,  \v \nabla \p(\v r, t)^{\dagger}\, \v\a\] \Big\} \ + \\ +  {1 \over 2} \[ \p(\v r, t)^{\dagger}\,, \b\,  \p(\v r, t)\] + j_{\mu}(\v r, t) A^{\mu}(\v r, t) \ ,
 \end{multline}
whence by the same arguments as for the current one obtains
\beq
\label{3.1h}
\ham_D (\v r, t)\, |vac\> = h_{D,VP}(\v r)\, |vac\>  \ ,
 \eeq
where
\begin{multline}
\label{3.1i}
h_{D,VP}(\vec{r})=\frac{1}{2}\(\sum\limits_{\e_{n}<\e_{F}} \e_n\,\p_{n}(\vec{r})^{\dagger}\p_{n}(\vec{r}) \ - \right. \\ \left. - \ \sum\limits_{\e_{n}\geqslant \e_{F}} \e_n\,\p_{n}(\vec{r})^{\dagger}\p_{n}(\vec{r}) \) \ .
\end{multline}
However, the VP-density $h_{D,VP}(\v r)$ itself turns out to be of no use in our approach, since it doesn't provide the direct  application of the WK method and so requires a more sophisticated techniques to deal with.

Upon integrating eq.(\ref{3.1i}) over the whole space we derive the expression for the VP-energy from the polarized Dirac sea
\beq
\label{3.1ii}
\E_{D,VP}=\frac{1}{2}\(\sum\limits_{\e_{n}<\e_{F}} \e_n  -   \sum\limits_{\e_{n}\geqslant \e_{F}} \e_n \) \ ,
\eeq
which again appears to be the eigenvalue of the total Dirac Hamiltonian acting on the vacuum state.

The origin of the Coulomb  term (\ref{1.2}) lies in the structure of the  EM-Hamiltonian
\beq
\label{3.68}
H_{EM} = {1\over 8 \pi}\,\int \! d{\v r}\, \[ \cvE^2+ \cvH^2 \] \ ,
\eeq
where
\beq
\begin{gathered}
\label{3.69}
\cvE(\v r,t)=-\v \nabla A^0(\v r,t) - {\pd \over \pd t} \v A (\v r,t)\, \ , \\ \cvH(\v r,t)=\v \nabla \times \v A(\v r,t)  \ .
\end{gathered}
\eeq
Upon splitting the electric field $\cvE(\v r,t)$ into the longitudinal and transversal components
\beq
\label{3.70}
\cvE(\v r,t)=\cvE_{||}(\v r,t)+\cvE_{\perp}(\v r,t) \ ,
\eeq
where $\cvE_{||}(\v r,t)=-\v \nabla A^0(\v r,t)$,  $\cvE_{\perp}(\v r,t)= - \pd \v A (\v r,t) / \pd t$, and integrating by parts, taking account of condition $\v \nabla \v A=0$, one gets  instead of expression (\ref{3.68})
\beq
\label{3.71}
H_{EM} = {1\over 8 \pi}\,\int \! d{\v r}\,  \cvE_{||}^2 + {1\over 8 \pi}\,\int \! d{\v r}\, \[  \cvE_{\perp}^2 + \cvH^2 \]\ ,
\eeq
where the first term can be easily transformed into the  operator-valued Coulomb interaction
\beq
\label{3.72}
\E_{Coul}=\frac{1}{2} \int \! d{\v r}\,d{\v r'}\, {j_{0}(\v r, t) \ j_{0}(\v r', t) \over |\v r - \v r'|} \ .
\eeq
Being  combined with the relation (\ref{3.1f}), the latter yields indeed the Coulomb term (\ref{1.2}) in the total VP-energy (\ref{1.3}).

The purpose of this work is to deal indeed with the expression (\ref{1.3}) in order to obtain  the correct  behavior of the VP-energy in the supercritical region for $Z>Z_{cr,1}$ with special attention to the contribution of the Coulomb term $\E_{C,VP}$.  In any case, however,  at first stage it would be instructive to consider the VP-density (\ref{3.1}) and vacuum shells formation within the WK framework.

\section{VP-density and vacuum shells formation in WK-techniques}

The essence of the WK techniques is the representation of the density (\ref{3.1}) in terms of contour integrals on the first sheet of the Riemann energy plane, containing the trace of the Green function of  the corresponding DC problem. In our case, the  Green function is defined via equation
\beq
\label{3.2}
\[-i \v{\a}\,\v{\nabla}_r +\b +V(r) -\e \]G(\vec{r},\vec{r}\,' ;\e)  =\d(\vec{r}-\vec{r}\,' ) \ .
\eeq

\begin{figure}
\center
\includegraphics[scale=0.20]{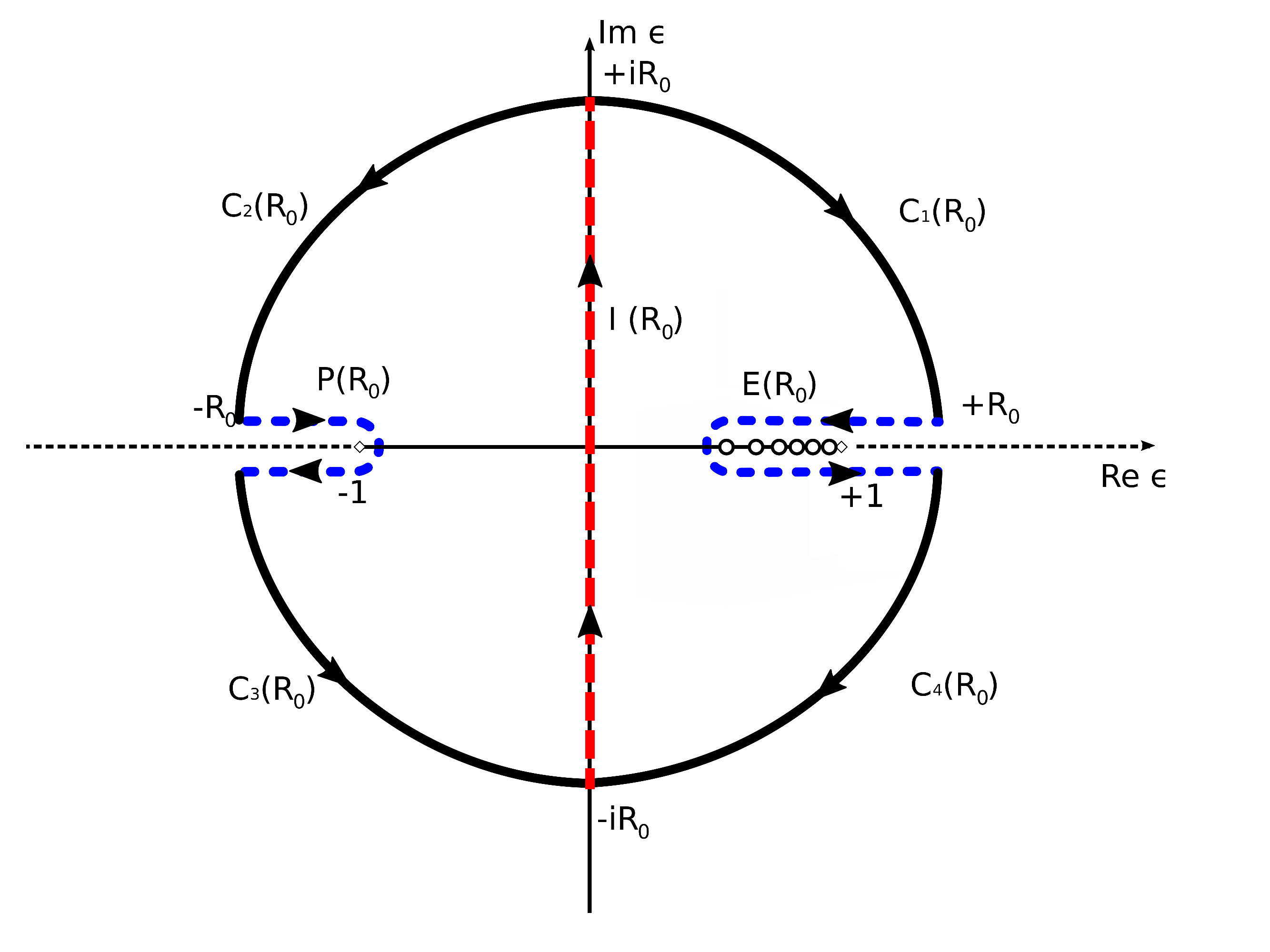} \\
\caption{\small (Color online) WK-contours in the complex energy plane, used for representation of the VP-charge density (\ref{3.1}) via contour integrals. The direction of contour integration is chosen in correspondence with (\ref{3.3}).}
\label{WK}
\end{figure}

The formal solution of (\ref{3.2}) reads
\beq
\label{3.3}
G(\vec{r},\vec{r}\ ';\e)=\sum\limits_{n}\frac{\p_{n}(\vec{r})\p_{n}(\vec{r}\ ')^{\dagger}}{\e_{n}-\e} \ .
\eeq
Following Ref.~\cite{Wichmann1956},  the density (\ref{3.1}) is expressed via the integrals along the contours  $P(R_0)$ and $E(R_0)$ on the first sheet of the complex energy surface  (Fig.\ref{WK})
\begin{multline}
\label{3.4}
\vr_{VP}(\vec{r}) = \\ = -\frac{|e|}{2} \lim_{R_0\rightarrow \infty}\( \frac{1}{2\pi i}\int\limits_{P(R_0)} \! d\e\, \mathrm{Tr}G(\vec{r},\vec{r'};\e)|_{\vec{r'}\rightarrow \v r} \ + \right. \\ \left.+ \ \frac{1}{2\pi i}\int\limits_{E(R_0)} \! d\e\, \mathrm{Tr}G(\vec{r},\vec{r'};\e)|_{\vec{r'}\rightarrow \v r} \) \ .
\end{multline}
Note that the Green function in this relation must be properly regularized to insure that the limit $\vec {r'} \to \v r$ exists and that the integrals over $d\e$ converge. This regularization is discussed below. On this stage, though, all expressions are to be understood to involve only regulated Green functions. One of the main consequences of the last convention is the uniform asymptotics of the integrands in (\ref{3.4}) on the large circle $|\e| \to \inf$ at least as $O(1/\e)$, which allows to transform the contours  $P(R_0)$ and $E(R_0)$ into the imaginary axis segment $I(R_0)$. Thereafter, upon taking the limit $R_0 \to \inf$, one obtains
\begin{multline}
\label{3.4a}
\vr_{VP}(\vec{r}) = \\ =|e| \[ \sum \limits_{-1 \leqslant \e_n < 0} |\p_n (\v r)|^2 +{1 \over 2\pi}\,  \int\limits_{-\inf}^{\inf} \! dy\, \mathrm{Tr}G(\vec{r},\vec{r'};iy)|_{\vec{r'}\rightarrow \v r}\]  \ ,
\end{multline}
where $\{\p_n(\v r)\}$ are the normalized eigenfunctions of negative discrete levels with $-1 \leqslant \e_n <0$ and here and henceforth
\beq\label{3.4b}
\p_{n}(\vec{r})^{\dagger}\p_{n}(\vec{r}) \equiv |\p_n (\v r)|^2 \ .
\eeq
Proceeding further, the Green function (\ref{3.3}) is represented as a partial series over $k=\pm(j+1/2)$ ~\cite{Wichmann1956,Gyulassy1975}
\beq
\label{3.12}
\mathrm{Tr}G(\vec{r},\vec{r'};\e)|_{\vec{r'}\rightarrow \v r} = \sum\limits_k { |k| \over 2 \pi}\,\mathrm{Tr}G_k(r,r';\e)|_{r'\rightarrow r} \ ,
\eeq
where the radial Green function  $G_k(r,r';\e)$ is defined via
\beq
\label{3.13}
\ham_k(r)\,G_k(r,r';\e)={\d (r-r') \over r r'} \ ,
\eeq
while the  radial DC hamiltonian takes the form
\beq
\label{3.13a}
\ham_k(r)=  \begin{pmatrix} V(r)+1-\e & -{1 \over r}\,{d \over dr}\,r - {k \over r} \\ {1 \over r}\,{d \over dr}\,r - {k \over r} & V(r)-1-\e  \end{pmatrix}  \ .
\eeq
 For the partial terms of $\vr_{VP}(r)$ one obtains
\begin{multline}
\label{3.14}
\vr_{VP,k}(r)  = {|e| |k| \over 2 \pi} \[ \sum \limits_{-1 \leqslant \e_{n,k} < 0} |\p_{n,k} (r)|^2 + \right. \\ \left. + {1 \over 2\pi}\,  \int\limits_{-\inf}^{\inf} \! dy\, \mathrm{Tr}G_k(r,r';iy)|_{r'\rightarrow r}\]  \ ,
\end{multline}
where $\p_{n,k} (r)$ are the normalized radial wave functions with eigenvalues $k$  and $\e_{n,k}$  of the corresponding radial DC problem. From (\ref{3.14}) by means of symmetry relations for $G_k(r,r';\e)$ from Ref.~\cite{Gyulassy1975}   for the sum $\vr_{VP,|k|}(r)$ of two partial VP-densities with opposite signs of $k$  one finds
\begin{multline}
\label{3.21}
\vr_{VP,|k|}(r)  =  {|e| |k| \over 2 \pi}\, \Big\{\sum \limits_{k=\pm |k|} \sum \limits_{-1 \leqslant \e_{n,k} < 0}  |\p_{n,k} (r)|^2  + \\ + {2 \over \pi}\,  \int\limits_{0}^{\inf} \! dy\, \mathrm{Re}\[\mathrm{Tr}G_{k}(r,r';iy)\]|_{r'\rightarrow r} \Big\} \ ,
\end{multline}
which is  by construction real and odd in $Z$ (in accordance with the Furry theorem).

The general result, obtained in Ref.~\cite{Gyulassy1975} within the expansion of $\vr_{VP}(r)$ in powers of  $Q$ (but with fixed $R$\,!)
 \beq
\label{3.22}
\vr_{VP}(\v r)  =  \sum \limits_{n=odd}\,Q^n\,\vr_{VP}^{(n)}(\v r) \ ,
\eeq
is that all the divergencies of $\vr_{VP}(\v r)$ originate  from the fermionic loop with two external photon lines, whereas  the next-to-leading orders of expansion in $Q$ are already free from divergencies (see also Ref.~\cite{Mohr1998} and refs. therein). This statement is valid for 1+1 and 2+1 D always, and for a spherically-symmetric external potential in the three-dimensional case.  In the  non-perturbative approach this statement has been verified for 1+1 D in Refs.~\cite{Davydov2017,Sveshnikov2017} and for 2+1 D in Refs.~\cite{Davydov2018a,*Davydov2018b,Sveshnikov2019a,*Sveshnikov2019b} via direct calculation. The latter can be also extended for the present 3+1 D case\,\footnote{We drop here all the intermediate steps, required for the explicit construction of radial Green functions $G_{k}(r,r';\e)$ for the potentials (\ref{1.5a},\ref{1.6}) and justification of the limit $r' \to r$. For these steps one needs to deal with  explicit solutions of the DC problem and corresponding Wronskians, which takes a lot of space and so will be considered separately.}. It should be mentioned, however, that in  the three-dimensional case the spherical symmetry of the Coulomb potential is crucial, since the above-mentioned statement can be justified first within the partial expansion of $\vr_{VP}(r)$ for each partial term  $\vr_{VP,|k|}(r)$, but not for the whole series at once.

 This circumstance has been discussed earlier in Ref.~\cite{Gyulassy1975}, where it was shown that  the main reason for such difference is that the properties of the partial $G_k$  are much better than for the whole  series. First, the limit $r' \to r$ exists for $G_k\(r, r'; \e\)$, while the limit $\vec {r'} \to \v r$ does not exist for the total $ G\(\v r,\v r'; \e\)$.   At the same time, for the first-order (linear in $Q$)  VP-density $\vr^{(1)}_{VP,|k|}(r)$  different results are obtained from  (\ref{3.21}) if the limit $r' \to r$ and the contour integral are interchanged.

 To the contrary,  the direct calculation of the total $\vr_{VP}^{(3)}(\v r)$ contribution by means of  (\ref{3.4a}) suffers from an ambiguity associated with the interchange of the limit $\vec {r'} \to \v r$ and integration over imaginary axis, but the calculation of the contribution from each partial $\vr^{(3)}_{VP,|k|}(r)$ is free from such ambiguities. This study of   $\vr_{VP}^{(3)}$-contribution suggests that for bounded potentials, regularization of $\vr_{VP}^{(3)}(\v r)$ is achieved  by calculating $\vr_{VP}^{(3)}(\v r)$ as a sum over the partial  contributions $\vr^{(3)}_{VP,|k|}(r)$ without any other manipulations.

So the renormalization of the VP-density (\ref{3.1}) is actually the same for all the three spatial dimensions and implies the following procedure.
First,  the linear in the external field terms in the expression  (\ref{3.4a}) should be extracted and replaced further by the properly renormalized  first-order perturbative density $\vr^{PT}_{VP}(r)$, evaluated for the same $R$. For these purposes let us introduce  the component  $\vr^{(3+)}_{VP,|k|}(r)$ of partial VP-density, which is defined in the next  way
\begin{multline}\label{3.25}
\vr_{VP,|k|}^{(3+)}(r)= \frac{|e||k|}{2 \pi}\,\Big\{ \sum\limits_{k=\pm|k|}\sum\limits_{-1\leqslant \e_{n,k}<0}|\p_{n,k}(r)|^2 \ +  \\  + \ \frac{2}{\pi} \int\limits_{0}^{\infty}d y\,\mathrm{Re}\[\tr G_{k}(r,r;iy)- \tr G^{(1)}_{k}(r;i y)\] \Big\} \ ,
\end{multline}
where $G^{(1)}_{k}(r;i y)$ is the  linear in $Q$ component of the partial Green function $G_{k}(r,r;i y)$ and so coincides with the first term of the Born series
\beq \label{3.26}
G^{(1)}_{k}=G^{(0)}_{k} (-V) G^{(0)}_{k} \ ,
\eeq
where $G_{k}^{(0)}$ is the free radial Green function with the same  $k$ and $\e$.
By construction $\vr^{(3+)}_{VP,|k|}(r)$ contains only the odd powers of $Q$, starting from $n=3$, and so is free of divergencies. At the same time, it is responsible for all the nonlinear effects, which are caused by   discrete levels  diving into the lower continuum.

The Born term (\ref{3.26}), which is required to extract the nonlinear component   from the total VP-density, for general Coulomb-like potential $V(r)$ of the type under question takes the form
 \begin{widetext}
\begin{multline}
\label{3.261}
 \tr G^{(1)}_{k}(r;i y) = - {1 \over r}\, \Bigg[(1-y^2)\,\(K^2_{k-1/2}(\g r)\,\int\limits_0^r \! dr'\,r'V(r')\,I^2_{k-1/2}(\g r') \ + \ K^2_{k+1/2}(\g r)\,\int\limits_0^r \! dr'\,r'V(r')\,I^2_{k+1/2}(\g r') +  \right. \\ \left.
 + I^2_{k-1/2}(\g r)\,\int\limits_r^{\inf} \! dr'\,r'V(r')\,K^2_{k-1/2}(\g r') \ + \ I^2_{k+1/2}(\g r)\,\int\limits_r^{\inf} \! dr'\,r'V(r')\,K^2_{k+1/2}(\g r')\)
+   \\  +
 (1+y^2)\,\(K^2_{k-1/2}(\g r)\,\int\limits_0^r \! dr'\,r'V(r')\,I^2_{k+1/2}(\g r') \ + \ K^2_{k+1/2}(\g r)\,\int\limits_0^r \! dr'\,r'V(r')\,I^2_{k-1/2}(\g r') +  \right. \\ \left.
 + I^2_{k-1/2}(\g r)\,\int\limits_r^{\inf} \! dr'\,r'V(r')\,K^2_{k+1/2}(\g r') \ + \ I^2_{k+1/2}(\g r)\,\int\limits_r^{\inf} \! dr'\,r'V(r')\,K^2_{k-1/2}(\g r')\)\Bigg] \ ,
\end{multline}
 \end{widetext}
where $I_\n(z)$ and $K_\n(z)$ being the Infeld and McDonald functions, while
\beq\label{3.262}
\g=\sqrt{1+y^2}  \ .
\eeq
In both cases (\ref{1.5a}) and (\ref{1.6}) the   Born term   can be   evaluated  in the analytical form,   but  the general answers are not given here explicitly because of their cumbersome form. The required explicit formulae for the potential (\ref{1.5a}) in the s-channel under question are listed in App.A.  The behavior of $\tr G^{(1)}_{k}(r;i y)$  for various asymptotical regimes of both arguments is considered in App.B. These results  predict some general features for the first-order VP-density $\vr^{(1)}_{VP,|k|}(r)$
\beq\label{3.263}
\vr^{(1)}_{VP,|k|}(r)= {|e| |k| \over \pi^2}\,    \int\limits_{0}^{\inf} \! dy\, \tr G^{(1)}_{k}(r;i y) \ ,
\eeq
which play an important role in what follows. Note that in $\vr^{(1)}_{VP,|k|}(r)$ there are no contributions from discrete levels, since they appear only in $\vr^{(3+)}_{VP,|k|}(r)$.

First, from the analysis of asymptotics of $\tr G^{(1)}_{k}(r;i y)$, presented in App.B, there follows that
\beq\label{3.264}
\vr^{(1)}_{VP,|k|}(r) \to   {C_1(k) \over r}   \ , \quad r \to 0 \ ,
\eeq
while for $r \to \inf$
\beq\label{3.265}
\vr^{(1)}_{VP,|k|}(r) \to   {C_2(k) \over r^3}   \ , \quad r \to \inf \ .
\eeq
with $C_1\,,C_2$ being some positive functions of the angular number $k$. An important and  general consequence of asymptotics (\ref{3.265}) is that the total first-order non-renormalized VP-charge
 \beq
\label{3.266}
Q^{(1)}_{VP}=\int\limits d {\v r}\,\vr^{(1)}_{VP}(\v r) = 4\,\pi\,\int\limits_0^\inf d {r}\,r^2\,\vr^{(1)}_{VP}(r)
\eeq
diverges.

\begin{figure}[ht!]
\subfigure[]{
		\includegraphics[width=1.02\columnwidth]{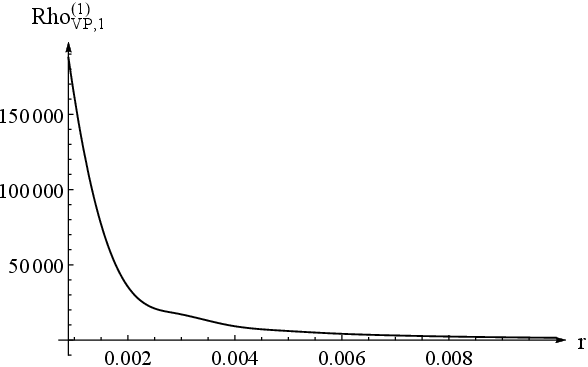}
}
\vfill
\subfigure[]{
		\includegraphics[width=0.95\columnwidth]{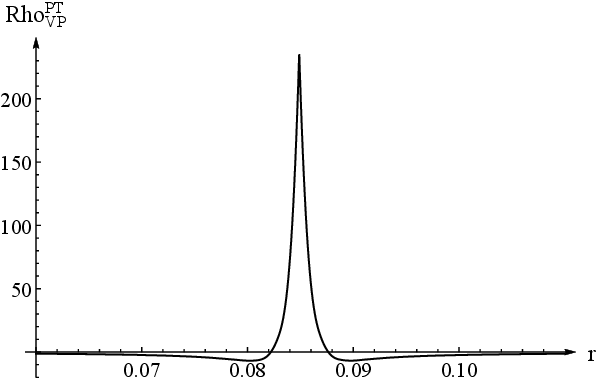}
}
\caption{The behavior of   $\vr^{(1)}_{VP,|k|}(r)$ in the $s$-channel, and $\vr^{PT}_{VP}(r)$, calculated in accordance with App.D, for $Z=200$ with $R=R_{1s} = 0.084888$. }
	\label{Rho1+PT_Z=200_R1s(r)}	
\end{figure}

Second, the behavior of $\vr^{(1)}_{VP,|k|}(r)$ as a function of $r$ is apparently different from  the perturbative properly renormalized (Uehling)  density $\vr^{PT}_{VP}(r)$. In particular, the calculation of $\vr^{(1)}_{VP,|k|}(r)$ by means of the formulae given in App.A, shows that it is always positive, while the Uehling one reveals a positive pick only in the vicinity of the Coulomb source location, being negative everywhere outside the latter (see  Fig.~\ref{Rho1+PT_Z=200_R1s(r)}). The detailed picture of the behavior of  $\vr^{(1)}_{VP,|k|}(r)$ on the whole range $0< r < \inf$ in the s-channel  is given in App.C.

This difference confirms once more that   the properly renormalized  VP-charge density $\vr^{ren}_{VP}(r)$ must be defined by the following expression\,\footnote{The convergence of the partial series in (\ref{3.27}) is shown explicitly for the point source in the original work by Wichmann and Kroll~\cite{Wichmann1956}, while accounting for the finite size of the source is discussed in detail in Ref.~\cite{Gyulassy1975}. For the present case it follows from convergence of the partial series for  $\E^{ren}_{VP}$, which is explicitly shown in~\cite{Grashin2022a}.}
\beq\label{3.27}
\vr^{ren}_{VP}(r)=\vr_{VP}^{PT}(r)+\sum\limits_{k=1}^{\inf}\vr_{VP,|k|}^{(3+)}(r) \ .
\eeq
The first-order perturbative, properly renormalized VP-density is obtained  in the next way ~\cite{Schwinger1949,*Schwinger1951,Itzykson1980, Greiner2012}
\beq
\label{2.2}
\vr^{PT}_{VP}(\vec{r})=-\frac{1}{4 \pi} \D A^{PT}_{VP,0}(\vec{r}) \ ,
\eeq
where
\beq\begin{gathered}
\label{2.3}
A^{PT}_{VP,0}(\vec{r})=\frac{1}{(2 \pi)^3} \int d \v q\,\, \mathrm{e}^{i \vec{q} \vec{r}}\, \Pi_{R}(-{\v q}^2)\,\widetilde{A}_{0}(\vec{q}) \ , \\
\widetilde{A}_{0}(\vec{q})=\int d \v r'\,\mathrm{e}^{-i \vec{q} \vec{r\,}' }\,A^{ext}_{0}(\vec{r}\,' ) \ .
\end{gathered}\eeq
The polarization function  $\Pi_R(q^2)$, which enters eq. (\ref{2.3}), is defined via general relation $\Pi_R^{\m\n}(q)=\(q^\m q^\n - g^{\m\n}q^2\)\Pi_R(q^2)$ and so is dimensionless. In the adiabatic case under consideration  $q^0=0$ and $\Pi_R(-{\v q}^2)$ takes the  form
\begin{multline}
\label{2.4}
\Pi_R(-{\v q}^2) = \\ = {2 \a \over \pi}\, \int \limits_0^1 \! d\b\,\b(1-\b)\,\ln\[1+\b(1-\b)\,{{\v q}^2 \over m^2-i\e}\] = \\ = {\a \over \pi}\, S\(|\v q|/m \) \ ,
\end{multline}
where
\begin{multline}\label{2.4a}
S(x)= -5/9 + 4/3 x^2 + (x^2- 2)\, \sqrt{x^2+4} \ \times \\ \times \ \ln \[ \(\sqrt{x^2+4}+x\) \Big/ \(\sqrt{x^2+4}-x\) \]/3x^3  \ ,
\end{multline}
with  the following IR-asymptotics
\beq
\label{2.4b}
S(x) \to x^2/15 -x^4/140  + O(x^6) \ , \quad x \to 0 \ .
\eeq

 The important feature of the renormalization procedure (\ref{3.27}) is that it provides vanishing of the  total renormalized VP-charge
\beq
\label{3.27a}
Q^{ren}_{VP}=\int\limits d {\v r}\,\vr^{ren}_{VP}(\v r) \ ,
\eeq
in the subcritical region $Z<Z_{cr,1}$. Actually, it is an  indication in favor of the assumption, that in presence of the external field, uniformly vanishing at the spatial infinity and without any specific boundary conditions and/or nontrivial topology of the field manifold, one should expect that in the region  $Z<Z_{cr,1}$ the correctly renormalized  VP-charge must vanish, while the VP-effects are able to produce only small distortions of its spatial distribution \cite{Greiner2012,Mohr1998}. It should be noted, however, that this is not a theorem, but just a plausible assumption, which in any concrete case should be verified via direct calculation.

In the present case the validity of this assumption can be shown in the following way. First, one finds that
\beq
\label{3.27b}
Q_{VP}^{PT} \equiv 0  \ .
\eeq
The relation (\ref{3.27b}) is nothing else but the direct consequence of the general QED-renormalization condition ${\Pi_R} (q^2) \sim  q^2$ for $q \to 0$, which is exactly confirmed by the asymptotics (\ref{2.4b}). More concretely, in the momentum space (up to factors like $2 \pi$ and general signs) the static equation for the potential $A_0(\v q)$, generated by the external charge density $\vr_{ext}(\v q)$, reads
\beq \label{2.5}
\({\v q}^2-{\tilde \Pi_R}(-{\v q}^2)\)A_0(\v q)=\vr_{ext}(\v q) \ ,
\eeq
where ${\tilde \Pi_R}(q^2)$ is the polarization operator in the standard form with dimension $[q^2]$, which is introduced via relation $\Pi_R^{\m\n}(q)=\(g^{\m\n}-q^\m q^\n/q^2\){\tilde \Pi_R}(q^2)$. Within PT one should propose that ${\tilde \Pi_R}(q)\ll q^2$, therefore  the potential $A_0(\v q)$ can be represented by the series
\beq \label{2.6}
 A_0(\v q)=A^{(0)}(\v q)+A^{(1)}(\v q)+\ldots.
 \eeq
 The classical part of the potential $A^{(0)}(\v q)$ is obtained from the equation
\beq \label{2.7}
{\v q}^2 A^{(0)}(\v q)=\vr_{ext}(\v q) \ ,
\eeq
and determines the external potential $A_0^{ext}(\v r)$, while its first quantum correction --- from the following one
\beq \label{2.8}
{\v q}^2 A^{(1)}(\v q)={\tilde \Pi_R}(-{\v q}^2)A^{(0)}(\v q)  \ .
\eeq
Up to the factor  $1/ 4 \pi$ the r.h.s. in (\ref{2.8}) is indeed the perturbative VP-density $\vr^{PT}_{VP}(\v q)$. Then in the coordinate representation (up to multipliers like $2\pi$) one gets
\begin{multline} \label{2.9}
\vr^{PT}_{VP}(\v r)=  \int \! d \v q \ \mathrm{e}^{i\v q \v r}\,{\tilde \Pi_R}(-{\v q}^2)\, A^{(0)}(\v q)= \ \\ \ = \int \! d \v r'\, {\tilde \Pi_R}(\v r-\v r') A^{ext}_{0}(\v r') \ .
\end{multline}
Upon integrating this expression over whole space  one obtains the total perturbative VP-charge $Q^{PT}_{VP}$ in the form
\beq \label{2.10}
Q^{PT}_{VP}=\int \! d \v q \ \d(\v q) \ {\tilde \Pi_R}(-{\v q}^2)\, A^{(0)}(\v q)   \ .
\eeq
Proceeding further by means of the  renormalization condition for ${\tilde \Pi_R}(q^2)$, one finds from  (\ref{2.10}) that  $Q^{PT}_{VP}=0$ under assumption, that in the momentum space  the singularity of the external potential $A^{ext}_{0}(\v q)$ for $|\v q| \to 0$ is weaker, than $O\(1/|\v q|^3\)$ in 2+1 D and than $O\(1/|\v q|^4\)$ in 3+1 D. For Coulomb-like potentials under consideration in 2+1 and 3+1 D the singularity of $A^{ext}_{0}(\v q)$ is $O\(1/| \vec q|\)$ and $O\(1/{\vec q}^2\)$, correspondingly.  In 1+1 D for the Coulomb potentials similar to (\ref{1.5a})   the singularity of $A^{ext}_{0}(\v q)$ is just a logarithmic one \cite{Davydov2017,Sveshnikov2017}, hence, $Q^{PT}_{VP}\equiv 0$ again.   Therefore for the  Coulomb potentials similar to (\ref{1.5a},\ref{1.6}) in all the three spatial dimensions there follows $Q^{PT}_{VP}\equiv 0$.

However, beyond the first-order PT and/or in the whole subcritical region $Z<Z_{cr,1}$, where in presence of negative discrete levels the dependence of $\vr_{VP}(\v r)$ on the external field cannot be described by the PT-series similar to (\ref{2.6}) any more, vanishing of $Q^{ren}_{VP}$ requires a sufficiently more detailed consideration.

\begin{figure}
\center
\includegraphics[scale=0.28]{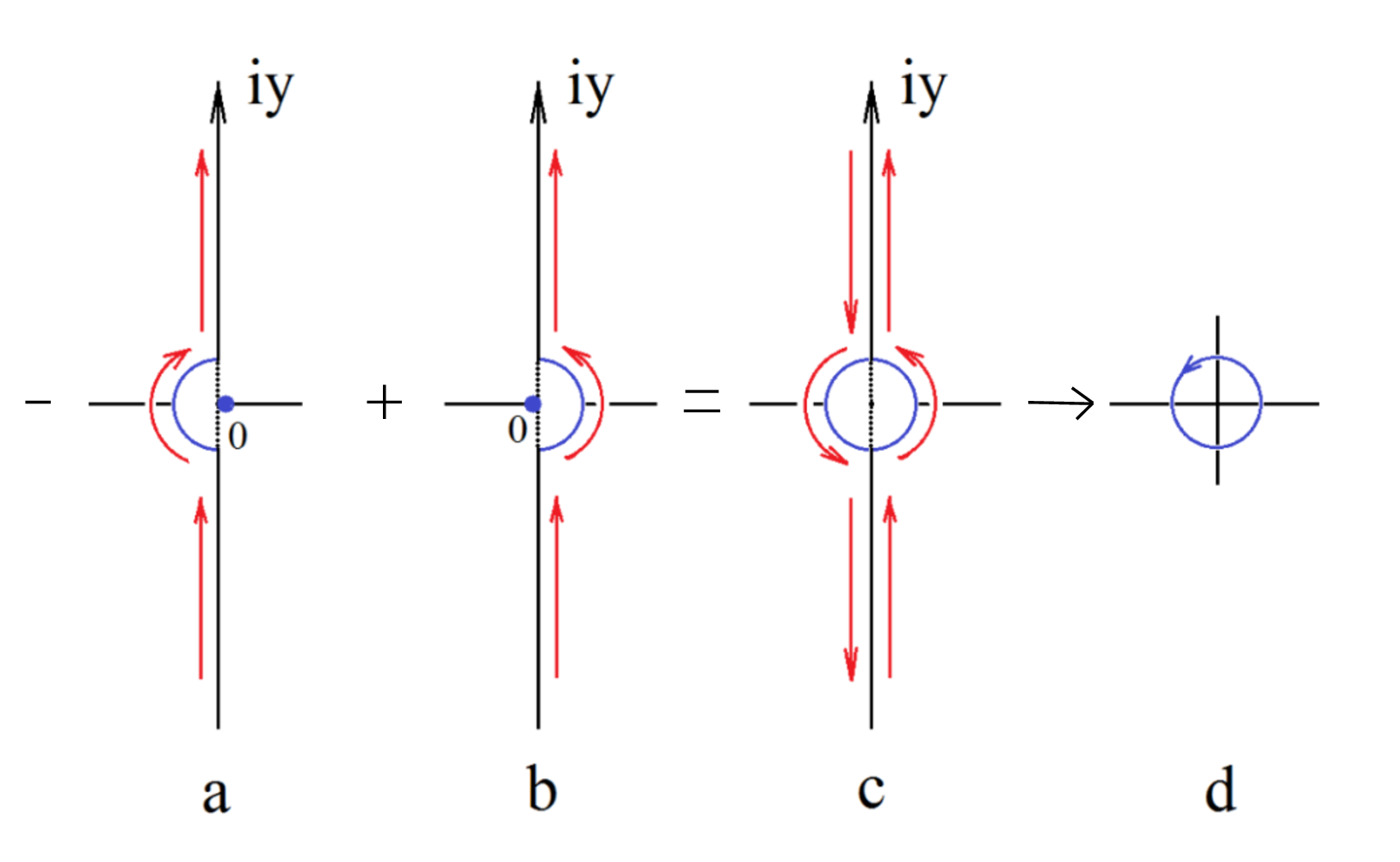} \\
\caption{\small (Color online) The picture of what happens with  the integral over the imaginary axis (\ref{2.11}), when the corresponding pole of the Green function moves along the real axis and  crosses zero point in the complex energy plane.}
\label{Residue}
\end{figure}

 In particular, the direct check confirms that the contribution of $\vr^{(3+)}_{VP}$ to  $Q^{ren}_{VP}$ for $Z<Z_{cr,1}$ also vanishes. In 1+1 D  this statement can be justified   purely analytically~\cite{Davydov2017}, while in 2+1 D due to complexity of  expressions, entering $\r^{(3+)}_{VP,|m_j|}(r)$, it requires a special combination of analytical and numerical methods (see Ref.~\cite{Davydov2018a}, App.B). For 3+1 D this combination is extended with minimal changes, since the structure of partial Green functions in two- and three-dimensional cases is actually the same up to additional factor $1/r$ and replacement $m_j \to k$. Moreover, it suffices to verify that  $Q_{VP}^{ren}=0$ not for the whole subcritical region, but only in absence of negative discrete levels. In presence of the latters,  vanishing of the  total VP-charge for $Z<Z_{cr,1}$ can be figured out by means of the  model-independent arguments, which are based on the initial  expression for the vacuum density (\ref{3.1}). It follows from (\ref{3.1}) that the change of $Q_{VP}^{ren}$ is possible only for  $Z>Z_{cr,1}$,  when the  discrete levels attain the lower continuum.   One of the possible correct ways to prove this statement is based on the rigorous analysis of the  behavior of the integral over the imaginary axis
 \beq \label{2.11}
 I_k(r)={1 \over 2\pi i}\,  \int\limits_{-i \inf}^{+i \inf} \! d\e\, \mathrm{Tr}G_k(r,r;\e) \ ,
 \eeq
 which enters the initial expression (\ref{3.14})  for $\vr_{VP,k}(r) $, under such infinitesimal variation of the external source, when the initially positive, infinitely close to zero level $\e_{n,k}$ becomes negative. Then the corresponding pole of the Green function undergoes an infinitesimal displacement along the real axis and also crosses the zero point, what yields the change in $I_k(r)$, equal to the residue at $\e=0$, namely
\beq \label{2.12}
 \D I_k(r)=\left. -|\p_{n,k}(r)|^2 \right|_{\e_{n,k}=0} \ .
 \eeq
For details see Fig.\ref{Residue} and take account of the definition of the Green function (\ref{3.3}).

So, by accounting for the degeneracy of levels in the spherically-symmetric case, the contribution from $I_k(r)$ to VP-charge loses  $2|e||k|$, whenever there appears the next negative discrete level  $\e_{n,k}$. Until this negative level exists, this loss of charge is compensated by the corresponding term, entering the sum  over negative discrete levels in (\ref{3.14}). Moreover, this term provides the continuity of the VP-density, when the level $\e_{n,k}$ crosses zero point. However, as soon as this level dives into the lower continuum,  the total VP-charge loses exactly $2|e||k|$.

 It should be specially mentioned that the latter effect is essentially non-perturbative and so completely included in $\vr_{VP}^{(3+)}$, while  $\vr_{VP}^{PT}$ is here out of work. So in the case under consideration one finds that upon renormalization the VP-charge turns out to be  non-vanishing only for $Z>Z_{cr,1}$ due to the non-perturbative effects, caused by diving of discrete levels into the lower continuum, in accordance with Refs.~\cite{Greiner1985a,Plunien1986,Mohr1998,Greiner2012,Rafelski2016}.

Thus, the behavior of the renormalized via eq.(\ref{3.27}) VP-density in the non-perturbative region turns out to be just the one, which should be expected from general assumptions about the structure of the electron-positron vacuum for $Z> Z_{cr}$. This result serves as a reasonable reference point for the whole renormalization procedure, and being combined with minimal subtraction prescription,  eliminates the inevitable arbitrariness in isolating the divergent part in the initial expression for $\vr_{VP}(\v r)$. This circumstance will be used further by renormalization of the VP-energy.

A more detailed picture of the changes in $\vr^{ren}_{VP}(r)$ for  $Z>Z_{cr,1}$ turns out to be quite similar to that considered in Refs.~\cite{Greiner1985a,Plunien1986,Greiner2012, Rafelski2016} by means of U.Fano approach to auto-ionization processes in atomic physics \cite{Greiner2012,Fano1961}. The main result is that, when the discrete level  $\p_{\n}(\v r)$ crosses the border of the lower continuum, the change in the   VP-charge density equals to
\beq\label{3.28}
\D\vr_{VP}^{ren}(\v r)=-|e|\times |\p_{\n} (\v r)|^2 \ .
\eeq
Here it  should be noted first that the original approach \cite{Fano1961} deals directly with the change in the  density of states $n(\v r)$ and so the change in the induced charge density (\ref{3.28}) is  just a consequence of the basic relation $\vr(\v r)= -|e| n(\v r)$.  Furthermore, such a jump in VP-charge density occurs for each diving level with its unique set of quantum numbers $\{\n\}$. So in the present case each discrete $2|k|$-degenerated level upon diving into the lower continuum yields the jump of total VP-charge  equal to  $(-2|e||k|)$. The other quantities, including the lepton number, should reveal a similar behavior. Furthermore, the Fano approach contains also a number of approximations. Actually, the expression (\ref{3.28}) is exact only in the  vicinity of the corresponding $Z_{cr}$, what  is shown by a number of concrete examples in Refs.~\cite{Davydov2017,*Sveshnikov2017,
Davydov2018a,Sveshnikov2019a,Voronina2019a}, and in the present context  in App.D for the potential  (\ref{1.5a}). Therefore, the most correct way to find $\vr_{VP}^{ren}(\v r)$ for the entire range of  $Z$ under study is provided by the expression (\ref{3.27}),  supplied with subsequent direct check of the expected integer value of  $Q_{VP}^{ren}$.

\section{VP-energy in the non-perturbative approach}

The most consistent way to explore the spontaneous  emission is to deal  with the total VP-energy $ \E_{VP} $, introduced in eq. (\ref{1.3}), since there are indeed the changes in $ \E_{VP} $, caused by discrete levels diving, which  are responsible for creation of vacuum positrons.   Following the general prescription for the Casimir energy calculations ~\cite{Itzykson1980,Plunien1986}, the natural choice of the reference point for $ \E_{VP} $ is the free case $A_{ext} = 0 $. The Coulomb term $ \E_{C,VP} $ is automatically consistent with such reference point. In the case of Dirac sea VP-energy it must be combined with the circumstance that, unlike the purely photonic Casimir effect, $ \E_{D,VP} $  includes also an infinite set of discrete Coulomb levels. To pick out  exclusively the interaction effects, it is therefore necessary to subtract in addition from each discrete level the mass of the free electron at rest.
In fact, such definition of $ \E_{D,VP} $ follows from the basic properties of QFT (see, e.g., the basic monographs~\cite{Bogoli'ubov1959, Bjorken1965}), which imply that the  in- and out- fields and states in any QFT-problem under question  should be subject of a special limiting procedure. The essence of the latter is that in any interacting system the asymptotically free in- and out- fields and states cannot be treated independently of dynamics. Actually, the free case should be defined   via pertinent  turning-off operation,  applied only to the final steps of calculation. In the case under consideration  it is  the limit $Q \to 0$. The specifics of the present problem is that for any infinitesimally small $Q$ the discrete spectrum of Coulomb states survives. So apart of continua, the free field limit in this case includes an infinite set of units, corresponding to discrete levels, concentrated in the infinitesimal vicinity of the condensation point, which must be subtracted in such a way, that preserves the physical content of the problem.

Thus, in the physically motivated form and in agreement with $\vr_{VP}(\v r)$, which is defined so that it automatically vanishes in the free case, the initial expression for the Dirac sea VP-energy should be written as
\begin{multline}
\label{3.29}
\E_{D,VP}=\1/2 \(\sum\limits_{\e_n<\e_F} \e_n-\sum\limits_{\e_n \geqslant \e_F} \e_n + \sum\limits_{-1\leqslant \e_n<1} \! 1
\)_A \ - \\ - \ \1/2 \(\sum\limits_{\e_n \leqslant -1} \e_n-\sum\limits_{\e_n
	\geqslant 1} \e_n \)_0 \ ,
\end{multline}
where the label A denotes the  non-vanishing external field $A_{ext}$, while the label 0 corresponds to   $A_{ext}=0$.  Defined in such a way,  the total VP-energy (\ref{1.3}) vanishes by turning off the external field, while by turning on it contains only the interaction effects. Therefore  the expansion of $ \E_{VP} $ in (even) powers of the external field starts from $ O\(Q^2\) $.

Now let us extract  from (\ref{3.29}) separately the contributions from the discrete and continuous spectra for each value of angular quantum number $k$, and afterwards use for the difference of integrals over the continuous spectrum $ (\int d {\v q}\, \sqrt{q^2+1})_A-(\int d {\v q}\,\sqrt{q^2+1})_0 $ the well-known technique, which represents this difference in the form of an integral of the elastic scattering phase $ \d_k(q)$~\cite{Sveshnikov2017,Rajaraman1982,Sveshnikov1991,Jaffe2004,Grashin2022a}.
 The final answer for $ \E_{D,VP}(Z,R) $, written as a partial series over angular number $k$, reads~\cite{Grashin2022a}
\beq
\label{3.30}
\E_{D,VP}(Z,R)=\sum\limits_{k=1}^{\inf} \E_{D,VP,k}(Z,R) \ ,
\eeq
where
\begin{multline}
\label{3.31}
\E_{D,VP,k}(Z,R)  = k\, \(\frac{1}{\pi} \int\limits_0^{\inf} \!   \  \frac{q \, dq }{\sqrt{q^2+1}} \ \d_{tot}(k,q) \ + \right. \\ \left. + \ \sum\limits_{\pm}\sum\limits_{-1 \leqslant \e_{n,\pm k}<1} \(1-\e_{n,\pm k}\)\) \ .
\end{multline}
In (\ref{3.31}) $ \d_{tot}(k,q) $ is the total phase shift for the given values of the wavenumber $q$ and angular number $\pm k$, including the  contributions from the scattering states from both  continua and  both parities for the radial DC problem with the  hamiltonian (\ref{3.13}). In the contribution of the discrete spectrum to $\E_{D,VP,k}(Z,R)$ the additional sum $\sum_{\pm}$ takes  also account of both parities.

Such approach to evaluation of $ \E_{D,VP}(Z,R)$ turns out to be quite effective, since for the external potentials of the type (\ref{1.5a},\ref{1.6})  each partial term in $ \E_{D,VP}(Z,R)$ turns out to be finite without any special regularization. First, $ \d_{tot}(k,q) $ behaves both in IR and UV-limits of the $q$-variable much better, than each of the scattering phase shifts separately.  Namely,  $ \d_{tot}(k,q) $ is finite for $ q \to 0 $ and behaves like  $ O(1/q^3) $ for $ q \to \inf $, hence, the phase integral in (\ref{3.31}) is always convergent. Moreover, $ \d_{tot}(k,q) $ is by construction an even function of the external field.  Second, in the contribution of bound states  to $ \E_{D,VP,k} $  the condensation point $ \e_{n,\pm k} \to 1 $ turns out to be regular, because  $ 1-\e_{n,\pm k} \sim O(1/n^2) $ for $ n \to \inf $. The latter circumstance permits to avoid intermediate cutoff of the Coulomb asymptotics of the external potential for $ r \to \inf $, what significantly simplifies all the subsequent calculations.

In the case of $ \E_{D,VP}(Z,R)$ the divergence of the theory shows up in divergence of the partial series (\ref{3.30}).  The latter problem can be solved along the lines of Refs.~\cite{ Davydov2018b,Sveshnikov2019b}, which deal with the similar  expansion for $ \E_{D,VP} $ in 2+1 D. In the present 3+1 D case due to a lot of technical details this topic is considered separately in Ref.~\cite{Grashin2022a}. The main result is that, as expected from general grounds discussed above in terms of $\vr_{VP}(\v r)$ and in accordance  with similar results in 1+1 and 2+1 D cases ~\cite{Davydov2017,*Sveshnikov2017,
Davydov2018a,Davydov2018b,Sveshnikov2019a,Sveshnikov2019b},  the partial series (\ref{3.30})  for $\E_{D,VP}(Z,R)$ diverges in the leading $O(Q^2)$-order. So it requires  regularization and subsequent renormalization, although each partial $\E_{D,VP,k}(Z,R)$  is finite. Moreover, the  divergence of the partial series (\ref{3.30}) is formally the same as in 3+1 QED  for the  fermionic loop with two external lines.

The need in the renormalization via fermionic loop follows also from the analysis of the properties of $\vr_{VP}$, which shows that without such UV-renormalization the integral VP-charge cannot acquire the expected integer value in units of $(-2|e|)$. In fact, the properties of $\vr_{VP}$ play here the role of a controller, which provides the implementation of the required physical conditions for a correct description of VP-effects beyond the scope of PT, since the latters cannot be tracked via direct evaluation of $\E_{D,VP}$ by means of the initial relations (\ref{1.3}),(\ref{3.29}).

Thus, in the complete analogy with the renormalization of VP-density  (\ref{3.27}), we should pass to the renormalized Dirac sea VP-energy by means of  relations
\beq
\label{3.58}
\E^{ren}_{D,VP}(Z,R) =\sum\limits_{k=1} \E^{ren}_{D,VP,k}(Z,R) \ ,
\eeq
where
\beq
\label{3.59}
\E^{ren}_{D,VP,k}(Z,R)=\E_{D,VP,k}(Z,R)+ \z_{D,k}(R) Z^2 \ ,
\eeq
while the renormalization coefficients $\z_{D,k}(R)$ are defined in the following way
\beq
\label{3.60}
\z_{D,k}(R) = \lim\limits_{Z_0 \to 0}  \[{\E_{D,VP}^{PT}(Z_0)\,\d_{k,1}-\E_{D,VP,k}(Z_0) \over Z_0^2}\]_{R} \ . \eeq
The  essence of relations  (\ref{3.58}-\ref{3.60}) is to remove (for fixed  $R$\,!) the divergent $O(Q^2)$-components from the non-renormalized partial terms  $ \E_{D,VP,k}(Z,R) $ in the series (\ref{3.30}) and  replace them further by renormalized via fermionic loop  perturbative contribution to the Dirac sea VP-energy $\E^{PT}_{D,VP}\,\d_{k,1}$. Such procedure provides simultaneously the convergence of the regulated this way partial series (\ref{3.58}) and the correct limit of $\E^{ren}_{D,VP}(Z,R)$  for $Q \to 0$ with fixed $R$.

Another way to achieve the renormalization prescription (\ref{3.58}-\ref{3.60}) can be based on the  Schwinger relation~\cite{Schwinger1949,*Schwinger1951,Sveshnikov2017,Plunien1986, Greiner2012} for the renormalized VP-quantities
\beq\label{3.61}
\d \E^{ren}_{D,VP}=\int \! \mathrm{d} \v r \ \vr^{ren}_{VP}(\v r)\, \d A_0^{ext}(\v r) +\d \E_N \ ,
\eeq
where $\E_N$ is responsible only for jumps in the VP-energy, caused by discrete levels crossing through the border of the lower continuum, and so is an essentially non-perturbative quantity, which  doesn't need any  renormalization. From (\ref{3.61}) there follows  that the Dirac sea VP-energy and VP-density must be renormalized in a similar way by means of the same subtraction procedure.

The relation (\ref{3.61}) can be represented also in the partial form
\beq\label{3.62}
\d \E^{ren}_{D,VP,k}=\int\limits_0^{\inf} \! r^2\, dr \ \vr^{ren}_{VP,k}(r)\, \d A_0^{ext}(r) +\d \E_{N,k} \ ,
\eeq
from which there follows that the  convergence of partial series for  VP-density implies the convergence of  partial series for  VP-energy and vice versa.  $\E_{N,k}$ is always finite and, moreover, for any finite $Z$ vanishes for $k \geq k_{max}(Z)$, therefore doesn't influence the convergence of the partial series.

The perturbative term $\E^{PT}_{D,VP}\,\d_{k,1}$ is obtained from the general first-order Schwinger relation~\cite{Schwinger1949,*Schwinger1951}
\beq
\label{2.1}
\E^{PT}_{D,VP}=\frac{1}{2} \int d \v r\, \vr^{PT}_{VP}(\vec{r})\,A_{0}^{ext}(\vec{r}) \ ,
\eeq
where $\vr^{PT}_{VP}(\vec{r})$ is defined in (\ref{2.2}-\ref{2.4a}).
From (\ref{2.1}) and (\ref{2.2},\ref{2.3}) one finds
\beq
\label{2.13}
\E^{PT}_{D,VP}=\frac{1}{64 \pi^4}\, \int \! d \v q \ {\v q}^2\, \Pi_R(-{\v q}^2)\,  \Big| \int \! d \v r\, \mathrm{e}^{i \vec{q} \vec{r}}\,A_{0}^{ext}(\vec{r}) \Big|^2  \ .
\eeq
It should be noted that, since the function  $S(x)$, defined in (\ref{2.4},\ref{2.4a}), is strictly positive, the perturbative VP-energy is also positive.

In the spherically-symmetric case  with $A_{0}^{ext}(\vec{r})=A_0(r)$  the perturbative VP-term belongs to the $s$-channel, which gives the factor $\d_{k,1}$, and
\begin{multline}
\label{2.14}
\E^{PT}_{D,VP}=\frac{1}{\pi}\, \int\limits_0^{\inf} \! d q \  q^4\, \Pi_R(-q^2) \ \times \\ \times \  \( \int\limits_0^{\inf} \! r^2\, d r\, j_0(q r)\, A_{0}(r) \)^2  \ ,
\end{multline}
 whence for the sphere there follows
\beq
\label{2.15}
\E^{PT}_{D,VP,\,sphere}={ Q^2  \over 2 \pi R}\, \int\limits_0^{\inf} \! {d q \over q} \ S(q/m) \ J_{1/2}^2(q R) \ ,
\eeq
while for the ball one obtains
\beq
\label{2.16}
\E^{PT}_{D,VP,\,ball}={ 9\,  Q^2  \over 2 \pi R^3}\, \int\limits_0^{\inf} \! {d q \over q^3} \ S(q/m) \ J_{3/2}^2(q R) \ .
\eeq
For $R$ close to $R_{min}(Z)$, under condition
\beq
\label{2.18}
\ln\( 1 / 2 m R \) \gg 1 ,
\eeq
the integrals (\ref{2.15},\ref{2.16}) can be  calculated analytically (see Ref.~\cite{Plunien1986} for details). In particular,
\beq\label{2.20}
\E^{PT}_{D,VP,\,sphere}={ Q^2  \over 3 \pi R}\,  \[ \ln\( {1 \over 2 m R}\) - \g_E + {1 \over 6}\]
\eeq
for the sphere and
\beq
\label{2.21}
\E^{PT}_{D,VP,\,ball}={2\, Q^2 \over 5 \pi R}\,  \[ \ln\( {1 \over 2 m R}\) - \g_E +  {1 \over 5} \]
\eeq
for the ball. In this case the ratio
\beq
\label{2.22}
\E^{PT}_{D,VP,\,ball}/\E^{PT}_{D,VP,\,sphere} \simeq 6/5 \
 \eeq
is just the same as for their classical electrostatic self-energies $ 3 Z^2 \a/5 R$ and $ Z^2 \a/2 R$.

However, for our purposes the  condition (\ref{2.18}) is too restrictive   and so the evaluation of $\E^{PT}_{D,VP}(R)$ is performed  numerically. By means of existing nowadays soft- and hardware this task does not pose any problems.

The renormalization of the Coulomb term $\E_{C,VP}(R)$ completely repeats the preceding ones. However, it by no means doesn't imply the direct replacement $\vr_{VP} (\v r) \to \vr^{ren}_{VP} (\v r)$ in the expression (\ref{1.2}). Rather,  it must follow the general lines of subtracting  (for fixed  $R$\,!) the  $O(Q^2)$-component from the non-renormalized expression  (\ref{1.2})  and  replacing it further by renormalized via fermionic loop  perturbative contribution to the Coulomb VP-energy $\E^{PT}_{C,VP}$. That means
\beq
\label{3.63}
\E^{ren}_{C,VP}(Z,R)=\E_{C,VP}(Z,R)+ \z_{C}(R) Z^2 \ ,
\eeq
while the renormalization coefficients $\z_{C}(R)$ should be defined now  as follows
\beq
\label{3.64}
\z_{C}(R) = \lim\limits_{Z_0 \to 0}  \[{\E_{C,VP}^{PT}(Z_0)-\E_{C,VP}(Z_0) \over Z_0^2}\]_{R} \ , \eeq
where
\beq\label{3.65}
\E_{C,VP}^{PT}=\frac{1}{2} \int \! d{\v r}\,d{\v r'}\, {\vr_{VP}^{PT}(\v r) \ \vr_{VP}^{PT}(\v r') \over |\v r - \v r'|} \ ,
\eeq
with $\vr_{VP}^{PT}(\v r)$ being defined in eqs. (\ref{2.2}-\ref{2.4a}).

Indeed such procedure deals with $\E_{C,VP}(R)$ in a way consistent with renormalization of $\vr_{VP}(\v r)$ and $\E_{D,VP}$ by means of the same subtraction procedure and provides  the correct limit of $\E^{ren}_{C,VP}(Z,R)$  for $Q \to 0$ with fixed $R$. A more explicit form of the expression (\ref{3.63}) can be also written as follows
\begin{multline}
\label{3.66}
\E^{ren}_{C,VP}(Z,R)= \\ \frac{1}{2} \int \! d{\v r}\,d{\v r'}\, {\vr_{VP}^{PT}(\v r)\, \vr_{VP}^{PT}(\v r') -\vr_{VP}^{(1)}(\v r)\, \vr_{VP}^{(1)}(\v r')\over |\v r - \v r'|} + \\ + \frac{1}{2} \int \! d{\v r}\,d{\v r'}\,{\vr_{VP}(\v r)\, \vr_{VP}(\v r') \over |\v r - \v r'|}   \ .
\end{multline}

The main point here is that in the supercritical region the Born term turns out to be much larger, than the perturbative one, which leads to a very peculiar effect of Coulomb supercriticality: Coulomb interaction between vacuum shells, being charged with the same (negative) sign, is attractive. This peculiar effect will be demonstrated by concrete calculations below.

\section{The results of calculations}

As in Ref.~\cite{Grashin2022a}, for greater clarity  of results  we restrict this presentation to the case of charged sphere (\ref{1.5a}) on the interval $0 < Z < 300$ with the numerical coefficient in eq. (\ref{1.8}) chosen as\,\footnote{With such a choice for a charged ball with $Z=170$ the lowest $1s_{1/2}$-level lies precisely at $\e_{1s}=-0.99999$. Furthermore, it is quite close to $1.23$, which is the most commonly used coefficient in heavy nuclei physics.}
\beq
\label{6.1}
R_{min}(Z)=1.228935\, (2.5\, Z)^{1/3} \ \hbox{fm} \ .
\eeq
On this interval of $Z$  the main contribution to VP-energy comes only from the s-channel, in which a non-zero number of discrete levels has already dived into the lower continuum. Therefore all the results, presented below, are given per two spin projections.

\begin{figure*}[ht!]
\subfigure[]{
		\includegraphics[width=1.5\columnwidth]{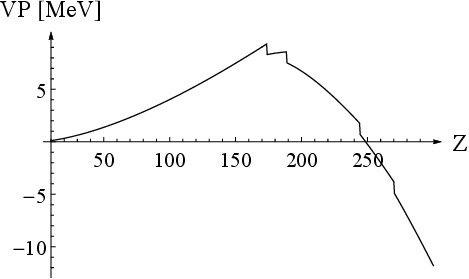}
}
\caption{$\E_{D,VP}^{ren}(Z)$ for $R=R_{min}(Z)$ in the range $10 < Z < 300$. }
	\label{EDVP(Z)}	
\end{figure*}

It would be instructive to start with  the  renormalized VP-energies of the Dirac sea $\E_{D,VP}^{ren}(Z)$ as functions of $Z$ with $R=R_{min}(Z)$, shown in Fig.\ref{EDVP(Z)}, and pertinent plots of $ \d_{tot}(1,q) $, shown in Figs.\ref{Phase(Z=300,R=Rmin)}  (for details of calculations see Ref.~\cite{Grashin2022a}).
\begin{figure*}[bt!]
\subfigure[]{
		\includegraphics[width=1\columnwidth]{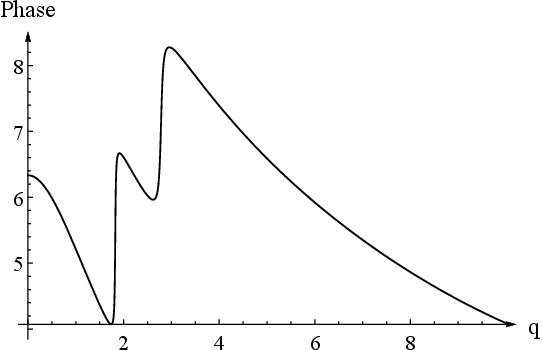}
}
\hfill
\subfigure[]{
		\includegraphics[width=1\columnwidth]{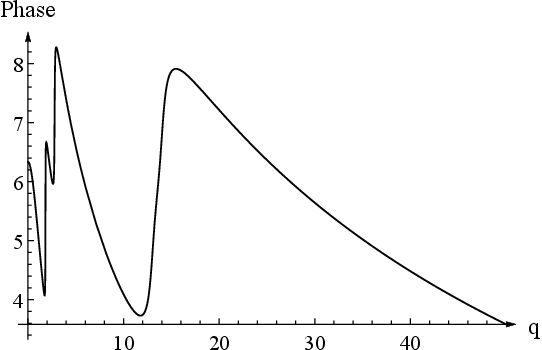}
}
\caption{$ \d_{tot}(1,q) $  for $Z=300$  and $R=R_{min}(Z)$ on different scales: (a)   $0 < q < 10$; (b) $0 < q <50 $. }
	\label{Phase(Z=300,R=Rmin)}\end{figure*}

The critical charges for both parities $(\pm k)$ are found in this case from equations
\begin{multline}\label{5.9}
2\, z_1\,K_{2 i \eta_k}\(\sqrt{8\,Q R}\)\,J_{\pm} \ \mp  \\ \mp \ \[\sqrt{2\,Q R}\,\(K_{1+2\, i \eta_k}\(\sqrt{8\,Q R}\) + K_{1-2 i \eta_k}\(\sqrt{8\,Q R}\)\)  \pm \right. \\ \left. \pm \ 2\,k\,K_{2 i \eta_k}\(\sqrt{8\,Q R}\)\]\,J_{\mp}=0 \ ,
\end{multline}
where
 \beq\label{4.29}
\eta_k=\sqrt{Q^2-k^2} \ ,
\eeq
and
\beq\label{5.10}
 z_1=\sqrt{Q^2-2\, Q R} \ , \quad J_{\pm}=J_{k \pm 1/2}(z_1) \  .
\eeq
The meaning of eqs. (\ref{5.9}-\ref{5.10}) is twofold. First,   by solving these eqs. with respect to $Z$ and $R=R_{min}(Z)$ one finds the standard set of critical charges $Z_{cr,i}$ for the case of charged sphere. Second, for fixed $Z$ one obtains the set of critical radii $R_{cr,i}(Z)$, for which the levels attain the threshold of the lower continuum.

\begin{figure*}[ht!]
\subfigure[]{
		\includegraphics[width=1.5\columnwidth]{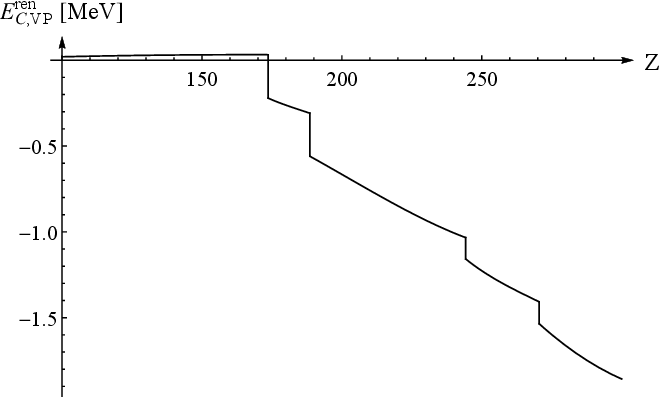}
}
\vfill
\subfigure[]{
		\includegraphics[width=1\columnwidth]{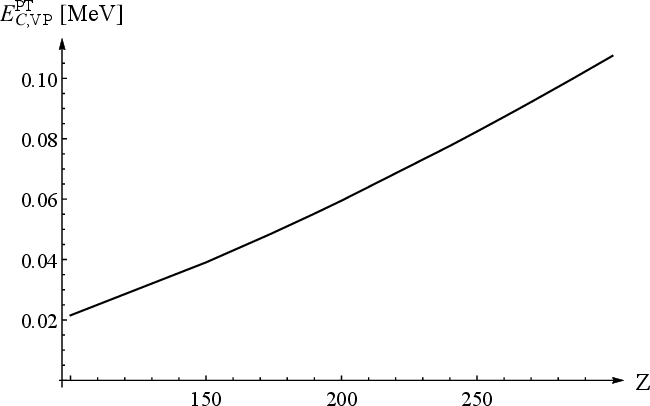}
}
\hfill
\subfigure[]{
		\includegraphics[width=1\columnwidth]{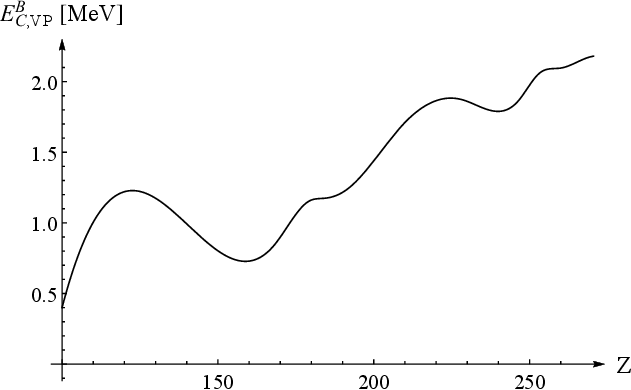}
}
\caption{Different aspects of the renormalized Coulomb term as a function of $Z$ and $R=R_{min}(Z)$ in the range $10 < Z < 300$: (a) $\E^{ren}_{C,VP}(Z)  $ ; (b) $ \E_{C,VP}^{PT}(Z)$ ; (c) $ \E_{C,VP}^{B}(Z)$. }
	\label{ECVP(Z)}
\end{figure*}
\begin{figure*}[ht!]
\subfigure[]{
		\includegraphics[width=1\columnwidth]{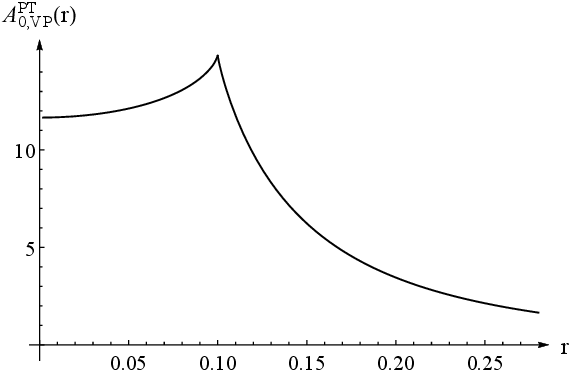}
}
\hfill
\subfigure[]{
		\includegraphics[width=1\columnwidth]{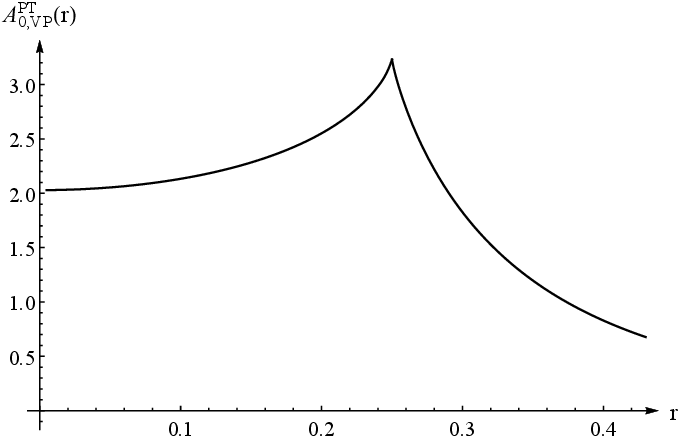}
}

\caption{The jumps of derivative in the Uehling potential $A^{PT}_{VP,0}(r)$ at $r=R$ for: (a) $Z=200\, , R=0.1$ ; (b) $Z=280\, , R=0.25$. The profiles of potentials are given in the bare form, as a direct result of corresponding integrations, omitting all the additional multipliers.}
	\label{Uehling_Z=200-280(r)}
\end{figure*}

\begin{figure*}[ht!]
\subfigure[]{
		\includegraphics[width=1.5\columnwidth]{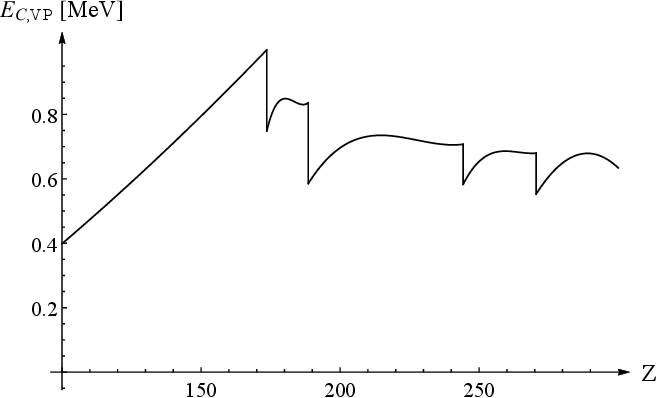}
}
\caption{The non-renormalized $ \E_{C,VP}(Z)$ as a function of $Z$ and $R=R_{min}(Z)$ on the range $10 < Z < 300$. }	
\label{ECoul(Z)}
\end{figure*}
For $Z=300$ there are four lowest levels in the $s$-channel, which  dive into the lower continuum. The dived levels  always group quite naturally into pairs, containing states of different parity. In this case these are the pairs $\{1s_{1/2}\, , 2p_{1/2}\}$ and $\{2s_{1/2}\, ,  3p_{1/2}\}$ with $Z_{cr,i} \simeq \{173.6\,, 188.5\}\, , \{244.3\, , 270.5\}$, correspondingly. So for $Z=300$ it is only the $s$-channel, where $\d_{tot}(1,q)$ undergoes corresponding resonant jumps by $\pi$. The pertinent curves of $\d_{tot}(1,q)$ are shown in Fig.\ref{Phase(Z=300,R=Rmin)}. In particular, the two first low-energy narrow jumps   correspond to resonances, which  are caused by diving of $2s_{1/2}$ and  $3p_{1/2}$, what happens quite close to $Z=300$. At the same time, the jumps caused by diving of $1s_{1/2}$ and  $2p_{1/2}$ have been already gradually smoothed and almost merged together, therefore look like one big $2\,\pi$-jump, which is already significantly shifted to the region of larger $q$. In the other channels with $k \geq 2$ there are no dived levels for such $Z$.

It should be remarked, however, that each $Z_{cr,i}$, given above, corresponds to the case of external Coulomb source with charge $Z=Z_{cr,i}$ and radius $R_i=R_{min}(Z_{cr,i})$. Hence, this is not exactly  the case under consideration with fixed $Z=300$  and $R=R_{min}(Z)$, rather it is  a qualitative picture of what happens with the dived levels in this case. They are absent in the discrete spectrum and show up as positronic resonances, the rough  disposition of which can be  understood as a result of diving of corresponding levels at $Z_{cr,i}$. But their exact positions on the $q$-axis can be found only via thorough restoration of the form of $ \d_{tot}(1,q)$.

 The contribution of $ \E_{C,VP}^{ren}(Z,R)$ is calculated as follows. In the $s$-channel the general expression (\ref{3.66}) reduces to
\begin{multline}
\label{3.67}
\E^{ren}_{C,VP}(Z,R)=(4 \pi)^2 \int_0^\inf \! r\, dr\,  \int_0^r \! r'^2\, dr'\,  \\ \[\vr_{VP}^{PT}(r)\,\vr_{VP}^{PT}(r') -\vr_{VP}^{(1)}(r)\, \vr_{VP}^{(1)}(r') + \vr_{VP}(r)\, \vr_{VP}(r')\]   \ ,
\end{multline}
where $\vr_{VP}^{PT}(r)\,,\vr_{VP}^{(1)}(r)\,, \vr_{VP}(r)$ are the perturbative renormalized, the Born one, defined in eq.(\ref{3.263}), and the total  VP-density,  calculated in the s-channel.

First, let us consider the  renormalized Coulomb interaction as a function of $Z$ with $R=R_{min}(Z)$. The pertinent plots of $\E^{ren}_{C,VP}(Z)  $ and of its main ingredients, in particular, the renormalized perturbative contribution $\E_{C,VP}^{PT}(Z)$ and the  Born term $ \E_{C,VP}^{B}(Z)$ are shown in Figs.\ref{ECVP(Z)}(a-c)  (for details of calculations see App. A-D).

As it was already claimed at the end of the preceding Section, the behavior of the renormalized $\E^{ren}_{C,VP}(Z)  $ corresponds to the picture, where before first level diving the Coulomb interaction between VP-densities is repulsive, but thereafter with each level diving at $Z_{cr,i}$, accompanied with appearance of vacuum shells, it undergoes negative jumps and so the Coulomb interaction between vacuum shells turns out to be attractive. This effect is a direct consequence of the renormalization prescription (\ref{3.63}-\ref{3.65}) and looks quite peculiar, since the shells are charged with the same sign. Moreover, in general the curve of $\E^{ren}_{C,VP}(Z)  $ looks quite similar to the one of the Dirac sea  with negative jumps at each $Z_{cr,i}$. The main difference is that the magnitude of $ \E_{D,VP}^{ren}(Z)$ is larger (and for larger $Z$ much larger) than that of the  Coulomb term, and with growing $Z$ goes down faster and faster in contrast to $\E^{ren}_{C,VP}(Z)$, which decreases much slowlier. Furthermore, the negative jumps in $ \E_{D,VP}^{ren}(Z)$ are always equal to $2 mc^2$, while those of $\E^{ren}_{C,VP}(Z)  $ depend on the structure of  wavefunctions $\p_n(\v r)$ at the lower threshold and so are not strictly fixed. Their general feature is that they decrease with growing $Z$, since there grows the number of nodes in $\p_n(\v r)$ at the lower threshold.

A remark should be also done concerning the calculation of the perturbative  term $\E_{C,VP}^{PT}(Z,R)$ for the sphere, since in this case the corresponding $\vr^{PT}_{VP}(r)$ reveals a $\d$-like singularity for $r=R$. Therefore the direct numerical evaluation of the corresponding term in the integral (\ref{3.67}) becomes difficult.  In this case  it will be more fruitful to deal directly with the Coulomb interaction written in terms of the longitudinal component of electric field $\cvE_{||}(\v r)$. For the potential (\ref{1.5a}) the Uehling potential $A^{PT}_{VP,0}(r)$ is continuous everywhere, but its derivative and so  $\cvE_{||,VP}^{PT}(r)$ reveal a finite jump at $r=R$. The pertinent plots of  $A^{PT}_{VP,0}(r)$ in the vicinity of the  discontinuity points are presented  in Figs.\ref{Uehling_Z=200-280(r)} for two quite representative values  $Z=200\,  , 280$. At the same time, for $r<R$ as well as for $r>R$ the radial electric field $\E_{||,VP}^{PT}(r)$ is well-defined. Hence, the perturbative contribution $\E_{C,VP}^{PT}(Z,R)$ can be quite consistently obtained  via
\beq
\label{3.72a}
 \E_{C,VP}^{PT}(Z,R)= {1\over 2}\,\[\int_0^R +\int_R^\inf\]  r^2\, dr\,  \(\E_{||,VP}^{PT}(r)\)^2 \ .
\eeq
Note also that in  Fig.\ref{ECVP(Z)}(b) the curve of $\E_{C,VP}^{PT}(Z)$ is not strictly quadratic in $Z$, since it is calculated for $R=R_{min}(Z)$.

In contrast to $\E_{C,VP}^{PT}(Z)$, which reveals a monotonic grow for increasing $Z$, the Born term  $ \E_{C,VP}^{B}(Z)$, which is built from the first-order $\vr_{VP}^{(1)}(r)$, is much more whimsical. First, due to asymptotics of $\vr_{VP}^{(1)}(r)$, given in (\ref{3.264},\ref{3.265}), the corresponding integral, which  defines the Born term in the expression (\ref{3.67}), is finite, and so the renormalization of the Coulomb term reduces to a finite subtraction. However, this subtraction is well motivated as a part of the whole renormalization procedure. Second, the magnitude of $ \E_{C,VP}^{B}(Z)$ is sufficiently larger, than of the other components of the Coulomb term. This is the reason, why  $\E^{ren}_{C,VP}(Z)  $ becomes negative in the supercritical region with $Z>Z_{cr,1}$. However, the most intriguing feature of $ \E_{C,VP}^{B}(Z)$ is that it shows a noticeable  reaction on the existence of critical charges. Compared to  $\E^{ren}_{C,VP}(Z)  $, this reaction is more soft, but, nevertheless, it is quite well pronounced. The reason is that unlike $\vr_{VP}^{PT}(r)$, which is constructed  as a purely perturbative quantity, $\vr_{VP}^{(1)}(r)$ is a part of the whole VP-density and therefore, at least partially, but reflects the information about the level diving, which is contained in the total $\vr_{VP}(r)$.

In addition, in Fig.\ref{ECoul(Z)} it is shown the non-renormalized Coulomb VP-energy $ \E_{C,VP}(Z, R_{min}(Z))$, which also turns out to be finite due to asymptotics (\ref{3.264},\ref{3.265}) of  $\vr^{(1)}_{VP,|k|}(r)$, contains the corresponding negative jumps at each $Z_{cr,i}$, but is strictly positive. Therefore without renormalization, which replaces $ \E_{C,VP}(Z,R)$ by $\E^{ren}_{C,VP}(Z,R)  $, the main effect of the Coulomb term (\ref{1.2}) will be the repulsion of VP-densities, which  would only worsen the situation with the spontaneous emission, which is already extremely bad. In contrast, $\E^{ren}_{C,VP}(Z,R)  $ works in a right direction and so makes the latter slightly better, but, unfortunately, cannot solve it completely.

\section{Spontaneous positron emission: what is wrong and what is possible}

Now --- having dealt with the general consequences of levels diving in terms of VP-density and VP-energy this way --- let us consider the total $ \E_{VP}^{ren}(Z,R)$ as a function of $R$.   This exercise turns out to be quite informative, since  this behavior of $\E_{VP}(Z,R)$  simulates in a quite reasonable way the non-perturbative VP-effects in slow heavy ion collisions, caused by  levels diving into the lower continuum, including the problem of spontaneous emission of positrons in the supercritical region $Z>Z_{cr,1}$\footnote{In this paper we give only a qualitative picture of what is wrong and what is possible concerning spontaneous emission in the supercritical region, based on the behavior of VP-density and VP-energy. More quantitative analysis with concrete calculations of emission rates and corresponding probabilities requires much more details including positron emission spectra, estimates of time intervals for emission, a set of additional plots and drawings, etc., and so will be presented separately.}.

First, let us mention that the general theory~\cite{Greiner1985a,Plunien1986,Greiner2012, Rafelski2016}, based on the  framework~\cite{Fano1961}, predicts that  after level diving there appears  a metastable state with lifetime $\sim 10^{-19}$ sec, which decays into  the spontaneous positron emission  accompanied with vacuum shells formation  according to  the Fano rule (\ref{3.28}).  An important point here is  that due to spherical symmetry of the source, during this process all the angular quantum numbers and parity of the dived level are preserved by the metastable state and further by positrons created. Furthermore, the spontaneous emission of positrons should be caused solely by VP-effects  without any other channels of energy transfer. The corresponding positron spectra have been calculated first in Refs.~\cite{Reinhardt1981,*Mueller1988,Ackad2008}, and explored quite recently with more details in Refs.~\cite{Popov2018,*Novak2018,*Maltsev2018,
Maltsev2019,*Maltsev2020}. These spectra demonstrate, in particular,  that  the emission of low-energy positrons should be strongly suppressed by the repulsive interaction with the nuclei, while at high energy the spectra fall off exponentially. Actually,  strong suppression of spontaneous emission at low energies is  the direct consequence of the  well-known result of Coulomb scattering  for the probability to find the scattering particles at zero relative distances
\beq
\label{6.4a}
|\P(0)|^2= { 2\, \pi \vk  \over v\, |\mathrm{e}^{2 \pi \vk}-1| } \ ,
\eeq
where $\vk=e_1 e_2/ \hbar v$ with $v$ being the relative velocity. In the case of repulsion for small $v$ the relation (\ref{6.4a}) implies
\beq
\label{6.4b}
|\P(0)|^2 \to  { 2 \pi e_1 e_2  \over \hbar\, v^2 }\, \mathrm{e}^{- 2 \pi e_1 e_2 /\hbar v}  \ ,
\eeq
and since the spontaneous positrons should be created in the vicinity of the Coulomb center,  there follows from (\ref{6.4b}) that one should expect strong suppression of emission at low energies.

\begin{figure*}[ht!]
\subfigure[]{
		\includegraphics[width=1.35\columnwidth]{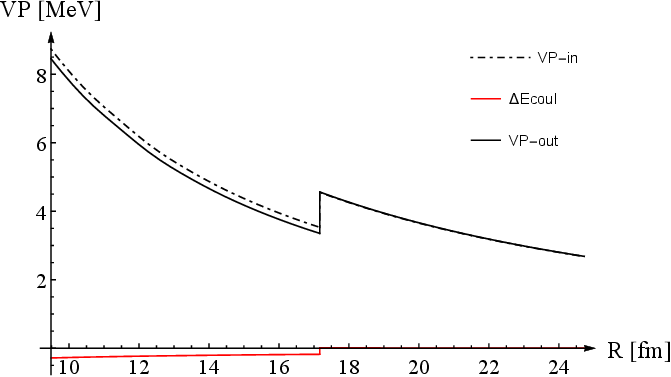}
}
\vfill
\subfigure[]{
		\includegraphics[width=1.35\columnwidth]{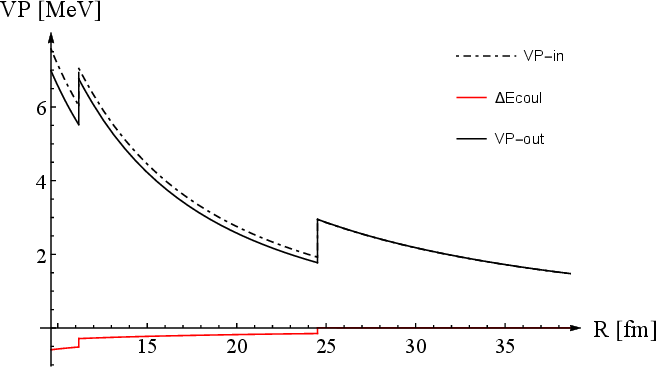}
}
\vfill
\subfigure[]{
		\includegraphics[width=1.35\columnwidth]{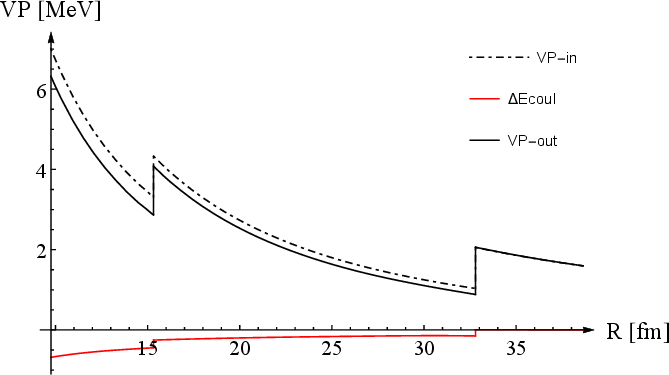}
}
\caption{(Color online). The total VP-energy (solid black line), built from contributions from  the Dirac sea $\E^{ren}_{D,VP}(Z,R)$ (dash-dotted black line) and Coulomb interaction  between VP-densities $\E^{ren}_{C,VP}(Z,R)  $ (red line) on the corresponding range of the Coulomb source size $R_{min}(Z) <R < R_{max} \simeq 30-40$ fm   and (a) $Z=184$; (b) $Z=192$; (c)  $Z=200$ .}
	\label{VP(Z=184-200,R)}	
\end{figure*}
\begin{figure*}[ht!]
\subfigure[]{
		\includegraphics[width=1.8\columnwidth]{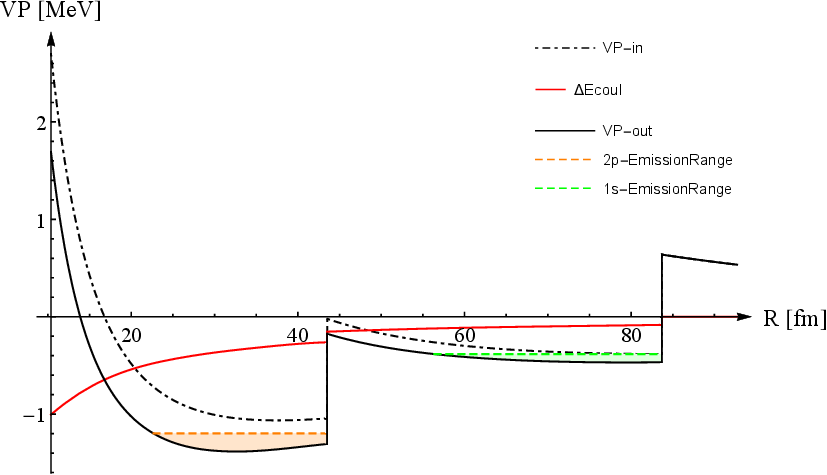}
}
\caption{(Color online).The total VP-energy (solid black line), built from contributions from  the Dirac sea  $\E^{ren}_{D,VP}(Z,R)$ (dash-dotted black line) and Coulomb interaction  between VP-densities  $\E^{ren}_{C,VP}(Z,R)  $ (red line) on the corresponding range of the Coulomb source size $R_{min}(Z) <R < R_{max} \simeq 80$ fm  and  $Z=240$. The areas of positive kinetic energy, allowed for spontaneous emission, are highlighted in color (1s---green\,, 2p---orange). }
	\label{VP(Z=240,R)}	
\end{figure*}
\begin{figure*}[ht!]
\subfigure[]{
		\includegraphics[width=1.8\columnwidth]{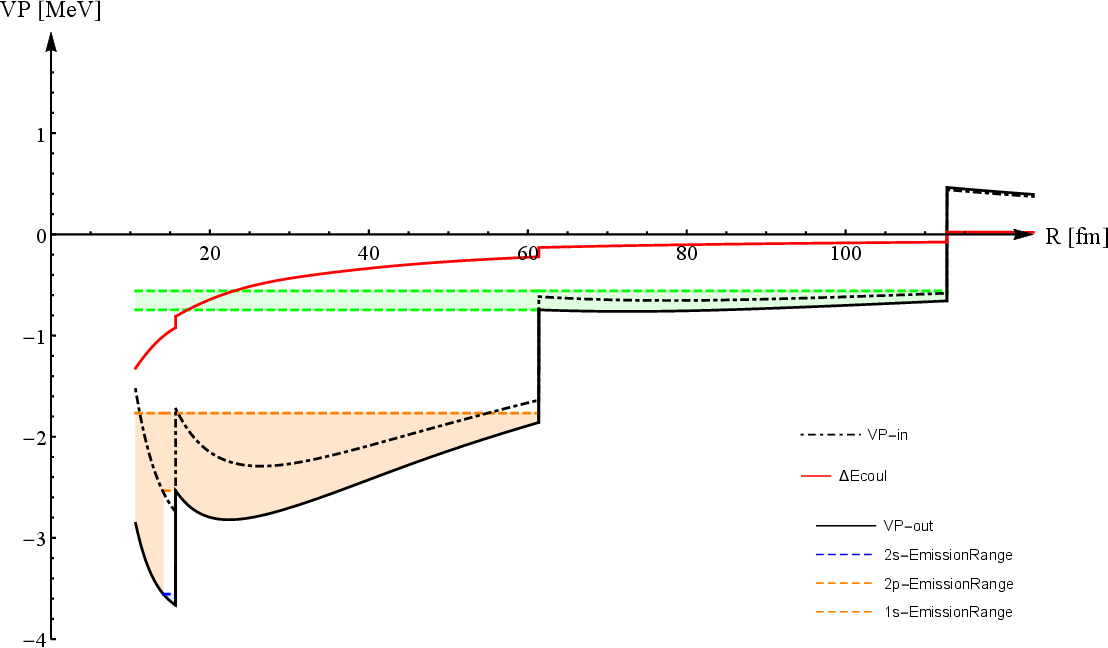}
}
\caption{(Color online).The total VP-energy (solid black line), built from contributions from  the Dirac sea  $\E^{ren}_{D,VP}(Z,R)$ (dash-dotted black line) and Coulomb interaction between VP-densities  $\E^{ren}_{C,VP}(Z,R)  $ (red line) on the corresponding range of the Coulomb source size $R_{min}(Z) <R < R_{max} \simeq 100$ fm  and  $Z=260$. The areas of positive kinetic energy, allowed for spontaneous emission, are highlighted in color (1s--green\,, 2p--orange\,, 2s--blue).}
	\label{VP(Z=260-m,R)}	
\end{figure*}
\begin{figure*}[ht!]
\subfigure[]{
		\includegraphics[width=1.8\columnwidth]{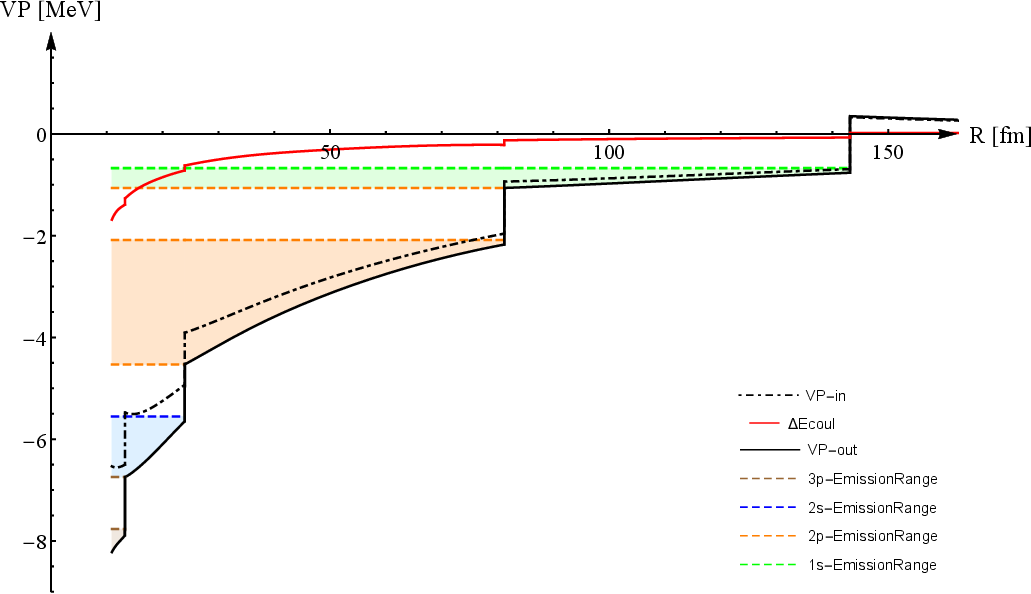}
}
\caption{(Color online). The total VP-energy (solid black line), built from contributions from  the Dirac sea  $\E^{ren}_{D,VP}(Z,R)$ (dash-dotted black line) and Coulomb interaction between VP-densities $\E^{ren}_{C,VP}(Z,R)  $  (red line) on the corresponding range of the Coulomb source size $R_{min}(Z) <R < R_{max} \simeq 150$ fm   and  $Z=280$. The areas of positive kinetic energy, allowed for spontaneous emission, are highlighted in color (1s--green\,, 2p--orange\,, 2s--blue, 3p--brown).}
	\label{VP(Z=280,R)}	
\end{figure*}
\begin{figure*}[ht!]
\subfigure[]{
		\includegraphics[width=1.8\columnwidth]{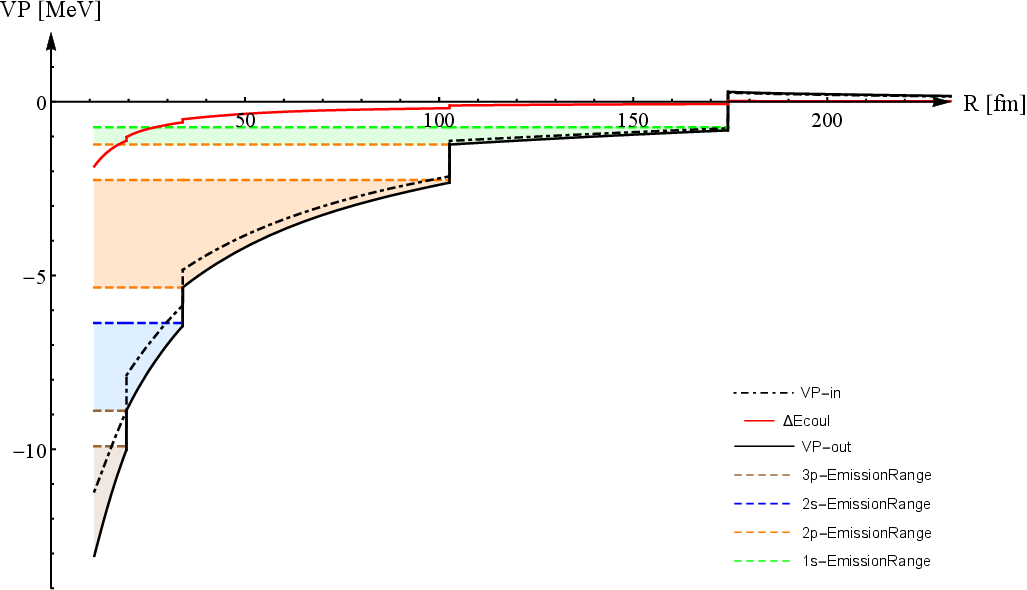}
}
\caption{(Color online). The total VP-energy (solid black line), built from contributions from  the Dirac sea  $\E^{ren}_{D,VP}(Z,R)$ (dash-dotted black line) and Coulomb interaction  between VP-densities   $\E^{ren}_{C,VP}(Z,R)  $ (red line) on the corresponding range of the Coulomb source size $R_{min}(Z) <R < R_{max} \simeq 200$ fm   and  $Z=300$. The areas of positive kinetic energy, allowed for spontaneous emission, are highlighted in color (1s--green\,, 2p--orange\,, 2s--blue, 3p--brown).}
	\label{VP(Z=300,R)}	
\end{figure*}
\begin{figure*}[ht!]
\subfigure[]{
		\includegraphics[width=1.5\columnwidth]{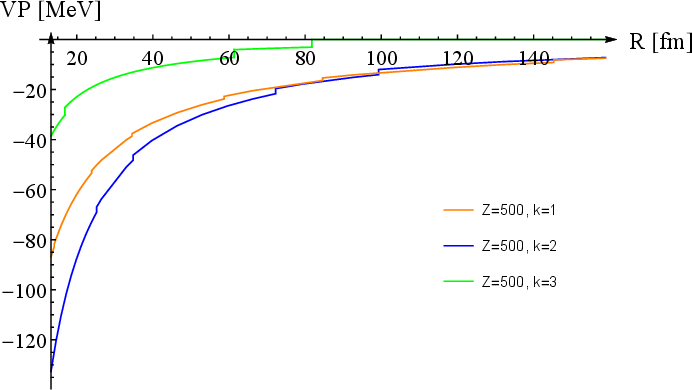}
}
\hfill
\subfigure[]{
		\includegraphics[width=1.5\columnwidth]{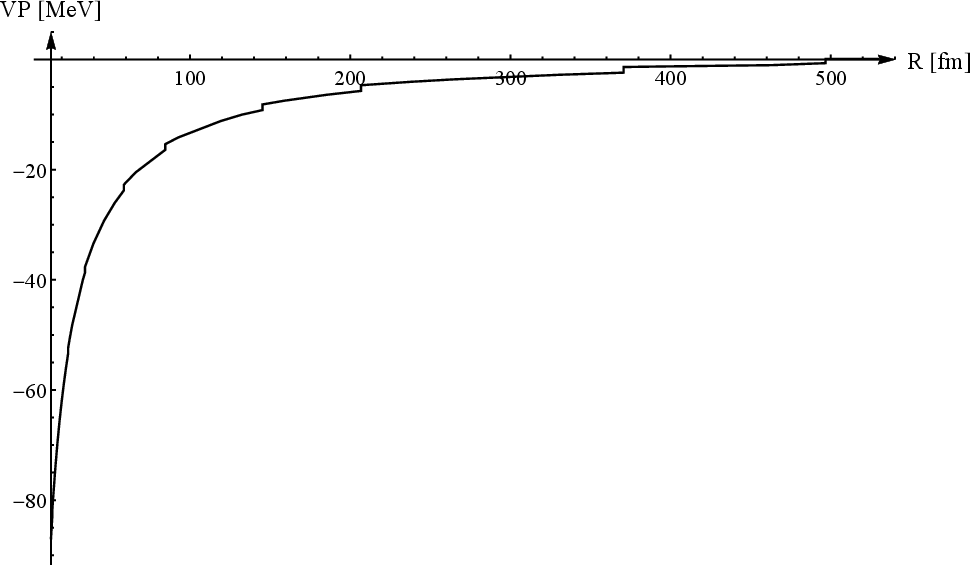}
}
\caption{(Color online) Partial  $ \E_{D,VP,k}^{ren}(R)$ for $Z=500$: (a) $k=1\,,2\,,3$ coupled together on the interval $0< R < 160$; (b) $k=1$ on a much larger interval $0< R < 550$ including all the jumps, caused by dived levels, and the asymptotical tail. }
	\label{VP500(R)}	
\end{figure*}

Now let us consider  $\E_{VP}^{ren}$ as a function of $R$ for a representative set of supercritical $Z$ from the range $184 \leq Z \leq 300$ for the $s$-channel with $k=1$. All the results, presented below, are given per two spin projections. The pertinent plots are shown in Figs.~\ref{VP(Z=184-200,R)}-\ref{VP(Z=300,R)}. Each curve is given  on the interval $R_{min}(Z) < R < R_{max}$, where $R_{max} \sim 1$ is chosen in such a way that it corresponds to the $O(1/R)$-asymptotics of $ \E_{VP}^{ren}(Z,R)$. The latter  appears as a result of competition between contributions from different components of $\E^{ren}_{D,VP}(Z,R)$ and $\E^{ren}_{C,VP}(Z,R)$, perturbative as well as non-perturbative. In the range $100<Z<300$ for the $s$-channel  this competition yields the  positive sign of asymptotics.  It should be noted also that the study of large $R$-asymptotics requires indeed the use of the exact integral representation of  $ \E_{D,VP}^{PT}(R)$, given in  (\ref{2.15}), instead of the approximate analytic expression  (\ref{2.20}), since in this region the condition (\ref{2.18}) apparently fails. At the same time, at the left end for $R=R_{min}(Z)$ the values of $ \E_{D,VP}^{ren}(Z,R)$ coincide with those from Fig.\ref{EDVP(Z)}.

In Figs.~\ref{VP(Z=184-200,R)} there are shown the curves of  $ \E_{VP}^{ren}(Z,R)$ for $Z=184\, , 192\, , 200$, which reveal the same basic property. Namely, their curvatures are strictly positive, and so they are bulge down. In particular, for $Z=184$ with $R_{min}(184)=0.02456 \simeq 9.5$ fm, there is only the lowest $1s_{1/2}$-level,  which dives into the lower continuum at $R_{cr,1}=0.0444 \simeq 17.2$ fm. The behavior of $ \E_{VP}^{ren}(Z,R)$ for $Z=192\,,200$ differs from $Z=184$ only in the number of dived levels and in the arrangement of curves  in accordance with their starting values for $R=R_{min}(Z)$. The curve $ \E_{VP}^{ren}(184,R)$ is the highest one with $ \E_{VP}^{ren}(184,R_{min}(184))\simeq 16.6\, \(\simeq 8.5\, \hbox{MeV}\)$, $\E_{VP}^{ren}(192,R)$ with $ \E_{VP}^{ren}(192,R_{min}(192))\simeq 13.6$ takes  the middle position, whereas the curve $ \E_{VP}^{ren}(200,R)$ is the lowest one with $ \E_{VP}^{ren}(200,R_{min}(200))\simeq 12.3$. Furthermore, for $Z=192\,,200$ there is already one pair of the lowest $s$-levels, namely $\{1s_{1/2}\, , 2p_{1/2}\}$, which dive into the lower continuum, now in reverse order compared to the dependence on $Z$ in Fig.\ref{EDVP(Z)}.  In particular, for $Z=200$ with increasing $R$ from $R_{min}$ to $R_{max}$, it is first the $2p_{1/2}$-level, which dives at $R_{cr,1}=0.044$, whereas  $1s_{1/2}$ dives much further at $R_{cr,2}=0.085$. Such a difference in $R_{cr,i}$ between $1s_{1/2}$ and $2p_{1/2}$  appears, because $Z_{cr,2} \simeq 188.5$ for $2p_{1/2}$ lies much closer to $Z=200$ than $Z_{cr,1} \simeq 173.6$ for $1s_{1/2}$.

$ \E_{VP}^{ren}(Z,R)$ for $Z=240\,,260$ differ from $Z=200$  in the number of dived levels, in the arrangement of curves  in accordance with their starting values for $R=R_{min}(Z)$ and, moreover, in the behavior of the derivatives $\nabla_R \E_{VP}^{ren}(Z,R)$. For $Z=240$ the derivative   $ \nabla_R \E_{VP}^{ren}(240,R)$ is negative almost everywhere, apart an interval between 20-40 fm, where it changes sign. However, this interval does not seriously influence the effects under question.   At the same time, $Z= 260$ turns out to be a  saddle point at which $ \nabla_R \E_{VP}^{ren}(260,R)$   changes sign on the already remarkably large interval $23 <R< 113$ (in fm) between the diving points  $1s_{1/2}$ and $2s_{1/2}$.

In contrast to $Z=200\, , 240\, , 260$, the behavior of $ \E_{VP}^{ren}(Z,R)$ for $Z=280\, , 300$ is principally different. First, the corresponding curves  lie remarkably lower, since the starting value for $R=R_{min}(Z)$ is in this case sufficiently more negative, namely $ \E_{VP}^{ren}(280,R_{min}(280))\simeq -16.0 \(\simeq -8.2\, \hbox{MeV}\)$,  $ \E_{VP}^{ren}(300,R_{min}(300))\simeq -25.6 \(\simeq -13.1\, \hbox{MeV}\)$. Second, they  reveal the curvature, which is opposite to the previous ones, and so they are bulge upwards. Moreover, with decreasing $R$ the decline of both curves into the negative range increases by each subsequent level diving. Indeed the latter property makes the spontaneous emission  a visible effect  for such $Z$.

 With further growth of $Z$ the decline  of $ \E_{D,VP}^{ren}(Z,R)$ into the negative range with decreasing $R$ becomes more and more pronounced, as it is clearly seen in Fig.\ref{VP500(R)}. $ \E_{C,VP}^{ren}(Z,R)$ shows up now only as a small correction. Moreover, there follows from Fig.\ref{VP500(R)}(a), that for $Z=500$ it is the $k=2$-channel, which yields  the main contribution to the total VP-energy decline. Note also, that for such $Z$ the asymptotics of $ \E_{D,VP}^{ren}(Z,R)$ for large $R$ becomes negative due to increasing role of nonlinear effects in VP-energy.

\begin{figure}
\center
\includegraphics[scale=
0.75]{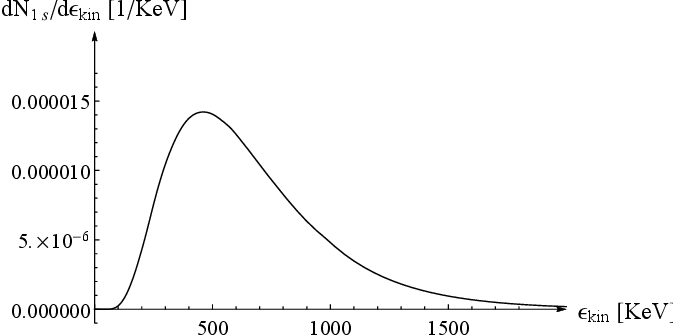} \\
\caption{\small Spontaneous emission distribution per additional kinetic energy in terms of average positron number $N_{1s}$ for $1s_{1/2}$ state at $Z=184$.}
\label{N1s_Z=184}
\end{figure}

Proceeding further, it would be instructive to deal with the process of  contraction, when $R$ varies from $R_{max}$ to $R_{min}$. This process simulates the approach of heavy ions to each other starting from large distances. In this case, the curves of $\E_{VP}(Z,R)$ in Figs.\ref{VP(Z=184-200,R)}  immediately show that for $Z=184-200$  the spontaneous emission is (almost) strictly forbidden. The reason is that although the rest mass of positrons can be created via negative jumps in  VP-energy at corresponding $R_{cr,i}$, which are exactly equal to $2 \times mc^2$ in accordance with two possible spin projections,   to create a real positron scattering state  it is not enough due to repulsion  between positrons and the Coulomb source.  To supply the emerging vacuum positrons with corresponding potential energy, which transforms further into the kinetic energy of scattering states, an additional decrease of  $\E_{VP}^{ren}(Z,R)$ just after levels diving point is required. However, the curves of $\E_{VP}(Z,R)$  for $Z=184-200$ do not share such option. On the whole interval  $R_{min} < R < R_{max}$,   with decreasing $R$ they reveal a  constant growth. The additional negative jumps   at critical points $R_{cr,i}$, caused by $\E_{C,VP}(Z,R)$, do not significantly improve the situation. In particular, for $Z=184$ the  $1s_{1/2}$-level dives at $R_{1s} \simeq 17.17$ fm with additional kinetic energy $\e_{kin}=186.8$ KeV, but the latter disappears with further contraction very rapidly, since  the whole region with $\e_{kin}>0$ occupies the interval $16.61 < R < 17.17$ (in fm). Under condition that the process of contraction stops at $R_{min}(Z)$, the average speed of contraction on this interval is $\simeq 0.07\, c$, while the corresponding transit time is estimated as $\D t_{1s} \simeq 2.6 \times 10^{-23}$ sec, what is two orders less than the resonance lifetime $\t_{res} \sim 10^{-19}$ sec \footnote{Remark that the time $\D t_{1s}$ is of pure kinematical origin and so has nothing to do with the width of spontaneous emission spectrum. As it follows from the general approach ~\cite{Greiner1985a,Plunien1986,Greiner2012, Rafelski2016, Fano1961}, the latter is coupled in the usual way with the resonance lifetime $\t_{res}$.}. In this case the spontaneous emission distribution, shown in Fig.\ref{N1s_Z=184}, should be given in terms of average positron number $N_{1s}$, rather than in terms of probability, since the latter cannot be properly normalized. In the next step, integrating the radial distribution
\beq
\label{6.4c}
dN_{\nu}/dR=dN_{\nu}/d\e_{kin}\times d\e_{kin}/dR \ ,
\eeq 
over the whole interval with $\e_{kin}>0$, for the total positron number one obtains $ N_{1s,tot} \simeq 0.0001$, what is much less than the conversion pairs background \footnote{As it was already mentioned above, the details of calculations will be presented in a separate paper.}.

\begin{figure}
\center
\includegraphics[scale=
0.75]{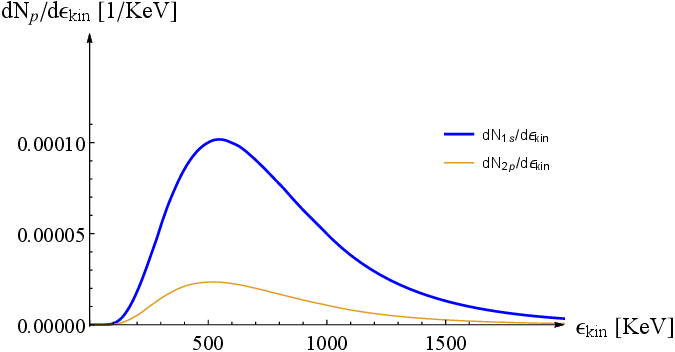} \\
\caption{\small (Color online). Spontaneous emission distribution per additional kinetic energy in terms of average positron numbers $N_{1s}$ for $1s_{1/2}$ and $N_{2p}$ for $2p_{1/2}$ states at $Z=200$.}
\label{Np_Z=200}
\end{figure}

For $Z=192-240$ one finds  actually the same situation. In particular, for $Z=200$ the  $1s_{1/2}$-level dives at $R_{1s} \simeq 32.8$ fm with additional kinetic energy $\e_{kin}=154.9$ KeV,   the region with $\e_{kin}>0$  occupies the interval $30.8 < R < 32.8 $ (in fm), the average speed of contraction on this interval is $\simeq 0.09\, c$, while the corresponding transit time equals to $\D t_{1s} \simeq 7.2\times 10^{-23}$ sec. For $2p_{1/2}$-level one gets correspondingly $R_{2p} \simeq 15.3$ fm with $\e_{kin}=189.4$ KeV, the interval $14.8 < R < 15.3 $ (in fm) for  the region with $\e_{kin}>0$, that is passed at the average speed  $\simeq 0.065\, c$ over time  $\D t_{2p} \simeq 2.6\times 10^{-23}$ sec. The spontaneous emission distribution is shown in Fig.\ref{Np_Z=200}. For the total positron number one obtains $ N_{1s,tot} \simeq 0.0002$ and $ N_{2p,tot} \simeq 0.00016$, that is still much less than the conversion pairs background.

For $Z=240$ the results are only slightly better. Although the areas of positive kinetic energy, where  the spontaneous emission is possible, are already quite large to be highlighted in color (1s--green\,, 2p--orange) in Fig.\ref{VP(Z=240,R)}, $\e_{kin}$	on these intervals do not exceed $85$ KeV for $1s_{1/2}$-level and $187$ KeV for $2p_{1/2}$-level. Corresponding transit times are of order $10^{-22}$ sec, while $ N_{2p,tot} \simeq 0.005$ and so is one order larger compared to $Z=200$, but   $N_{1s,tot}  \simeq   6 \times 10^{-6}$ is even lower, that for $Z=184-200$. This happens because the $1s$-diving point is now quite large $R_{1s}=83.7$ fm \footnote{Such a large value of $R_{1s}$ in this case should not be misleading, since the total $Z=240$ cannot be realized in two heavy ions collision. One needs at least four ions, moving along the tetrahedron carbon bonds. The last configuration is quite  close to a sphere even for such $R$.}. Meanwhile, the additional negative jump caused by $\E^{ren}_{C,VP}(Z,R)$ decreases with increasing $R$, and so $\e_{kin}$ for $1s_{1/2}$-level is restricted  by $85$ KeV from above. Therefore the whole interval, allowed for spontaneous emission of $1s_{1/2}$ positrons, belongs to the left asymptotics of emission distribution with vanishing probability according to relation (\ref{6.4b}).

The approximate starting point   of spontaneous emission is presented by the curve $\E_{VP}(260,R)$, shown in Fig.\ref{VP(Z=260-m,R)}. In this case $1s_{1/2}$ and $2s_{1/2}$ levels reveal actually the same features with very low emission rates as for $Z=240$ and $Z=184-200$. However,  spontaneous emission of  $2p_{1/2}$-state takes place now in the whole region between $R_{min}(260)=10.64$ fm and  $R_{2p}=61.4$ fm, and so proceeds with  an appreciable probability. First, it is the interval $15.7 < R < 61,4$ (in fm), where by decreasing $R$ the curve $\E_{VP}(260,R)$  first goes down, providing sufficient growth of $\e_{kin}$ up to $\simeq 1056$ KeV at $R \simeq 22.5$ fm. Second, there appears an additional  interval $10.64< R < 14.1$ fm between $R_{min}$ and the endpoint of positron emission in the $2s_{1/2}$-state. In this region $\e_{kin}$ exceeds a quite large value $\simeq 1786$ KeV, which provides the emission probability $\simeq 0.98$. Moreover, the total transit time from $R_{2p}$ to $R_{min}$ is estimated as $\simeq 2 \times 10^{-19}$ sec and so is quite close to $\t_{res}$. So for $Z=260$ the spontaneous emission of $2p_{1/2}$ positron states becomes an accomplished fact.

Actually,  the start-up for spontaneous emission should be estimated as $Z^{\ast} \simeq 250-260$, since it will not happen until $\E_{VP}(Z,R_{min}(Z))$ becomes negative. But there follows from Fig.\ref{EDVP(Z)} that the latter cannot occur before $Z$ exceeds $250$. Moreover, this estimate for $Z^{\ast} $ turns out to be quite insensitive to the concrete model of the Coulomb source. The main reason is that this estimate is closely related to the condition $\E_{VP}(Z,R_{min}(Z)) \simeq 0$. At this point, the difference between the sphere and ball charge configurations is small, since for $R=R_{min}(Z)$ it is reliably confirmed in Ref.~\cite{Grashin2022a} that their VP-energies are related via ratio (\ref{2.22}). So in the positive range the curve of the ball VP-energy lies higher than the corresponding curve for the sphere, in the negative range they interchange,  whereas their intersection takes place indeed at the point $\E_{VP}(Z,R_{min}(Z)) \simeq 0$, i.e. for the same $Z$ up to corrections in inverse powers of $\ln\( 1 / 2 m R_{min}(Z)\)$.  At the same time, in slow heavy ion collisions it is well-known that the VP-effects at short internuclear distances, achieved in the monopole approximation, are in rather good agreement with exact two-center ones~\cite{Reinhardt1981,*Mueller1988,Tupitsyn2010,Maltsev2019,*Maltsev2020},  and lie always in between those for sphere and ball upon adjusting properly the coefficient in the relation (\ref{1.8}).

However, for $Z \simeq Z^{\ast}$ the intensity of spontaneous emission cannot be high enough against the conversion pairs background. In particular, for $Z=260$ there are only $2p_{1/2}$- positrons, which are emitted after each diving of $2p_{1/2}$-level. Actually, the positron emission becomes a visible effect starting from $Z \sim 280-300$ and with further growth of $Z$ its intensity increases very rapidly, which opens up the possibility for unambiguous detection for such $Z$. It happens because  the behavior of $ \E_{VP}^{ren}(280,300,R)$   is opposite to the previous ones (compare the appearance of curves in Figs.~\ref{VP(Z=184-200,R)}--\ref{VP(Z=260-m,R)} and Figs.~\ref{VP(Z=280,R)},\ref{VP(Z=300,R)}). As a result, with decreasing $R$ the decay rate  of $ \E_{VP}^{ren}(280,300,R)$ into the negative range increases by each subsequent level diving and so supplies the positrons just born with sufficient amount of energy and  transit time  for emission in the vicinity of the Coulomb source. And it is indeed the latter property, which makes the spontaneous emission  a quite visible effect  for such $Z$. With further growth of $Z$,  by each subsequent levels diving  the decline  of $ \E_{VP}^{ren}(Z,R)$  into the negative range  with  decreasing $R$ becomes more and more pronounced, as it is clearly seen in Fig.\ref{VP500(R)}.  As an additional factor of growth here is that   the  channels with $k >1 $ contribute with corresponding multiplicity factors. So  the total rate of positron emission becomes significantly more high.

\section{Concluding remarks}

First, it should be noted that the answer $Z^{\ast} \simeq 250-260$ for the start-up of spontaneous emission, given above, is just an estimate from below. Actually, $Z^{\ast}$ should be even larger, since this answer corresponds to a strongly underestimated  size of the Coulomb source via relation (\ref{6.1}) in comparison with the realistic heavy ions supercritical configuration. Moreover, there exists at least one correction to $ \E_{VP}^{ren}(Z,R)$, which leads to a positive shift of $Z^{\ast}$.  It is the self-energy (SE) contribution  due to virtual photon exchange, which in the context of essentially nonlinear VP-effects of super-criticality, caused by  fermionic loop,  is usually dropped. Near the lower continuum  SE shows up also as a perturbative correction   and so cannot seriously alter the results, presented above~\cite{Roenko2018}. Nevertheless,    SE is always positive and so leads to the same consequences for  $Z^{\ast}$. So the realistic estimate for $Z^{\ast}$ turns out to be even higher than  $ \simeq 250-260$. In any case, however, such $Z^{\ast}$ lies far beyond  the interval $170 \leqslant Z \leqslant 192 $, which is nowadays the main region of theoretical and experimental activity in heavy ion collisions aimed at the study of such VP-effects ~\cite{Popov2018,Maltsev2019,FAIR2009,Ter2015,MA2017169}.
The negative result of early investigations at GSI~\cite{Mueller1994} can be  at least partially explained by the last circumstance\,\footnote{We drop here any questions concerning the time of level diving during heavy ion collisions. For more details on this topic see, e.g., Refs.~\cite{Rafelski2016, Greiner2012,Ruffini2010}.}.

Note also that there follows from results of Sect.VI that for  $Z=184-240$ the probability of positron emission after corresponding level diving is very low. In this case the dived level penetrates into the lower continuum as an empty one, creating only the change in the density of states $n(\v r)$  according to the general approach \cite{Greiner2012,Fano1961}. The emerging vacuum  shell  remains uncharged and so cannot yield any additional negative jump in $\E^{ren}_{C,VP}(Z,R) $ at corresponding diving point. $\E^{ren}_{C,VP}(Z,R) $  contains in this case $\E^{PT}_{C,VP}(Z,R) $ as the main contribution, which is positive and very small compared with $\E^{ren}_{D,VP}(Z,R)$. In Fig.\ref{ECVP(Z)}(b) it is shown as a function of $Z$ for $R=R_{min}(Z)$, while for growing $R$ and fixed $Z$ it vanishes as $O(1/R)$. Therefore, the total  $\E^{ren}_{VP}(Z,R)$ coincides in this case  with the bare Dirac sea polarization energy $\E^{ren}_{D,VP}(Z,R)$, shown in Figs.\,\ref{VP(Z=184-200,R)}-\ref{VP(Z=300,R)} as a dash-dotted black line, up to very small positive correction. This is why we can freely ignore in eq.(\ref{3.2}) an addition to  the external  potential $V(r)$, caused by VP-charge density, which appears  in the Coulomb gauge
\beq
\label{3.2i}
\D V(\v r)=e \int \! d{\v r'}\, {\vr_{VP}(\v r') \over |\v r - \v r'|} \ .
\eeq

       It should be also mentioned that  $Z^{\ast} \simeq 210 $  has already arisen by treating the spontaneous emission as a specific lepton pair creation~\cite{Grashin2022a,Krasnov2022,*Krasnov2022a}. The latter estimate  is significantly softer, but has been  obtained on the basis of more qualitative  arguments. The present one, however, is achieved in a quite rigorous way via direct evaluation of $ \E_{VP}^{ren}(Z,R)$ by means of well-founded and transparent approach.

Second, in collisions with spontaneous emission the emitted positrons carry away the lepton number equal to $(-1)\times$their total number. Hence, the   corresponding amount of positive lepton number  should be left to the  VP-density, concentrated in vacuum shells. Otherwise, either the lepton number conservation  in such a processes must be broken, or the positron emission  prohibited. Therefore such collision should proceed via a specific intermediate state, which contains a spatially extended lepton number VP-density. Moreover,  it should be a real intermediate state, which appears with the first emitted positron and ceases  to exist with the last created valence electron. For slow heavy ion collisions its lifetime can be estimated as $\sim 10^{-19}-10^{-21}$ sec, and during this time the lepton number VP-density should exist as a real and, in principle,  measurable extended quantity.
So any reliable answer concerning the spontaneous positron emission --- either positive or negative --- is important for our understanding of the nature of the lepton number, since so far leptons show up as point-like particles with no indications on existence of any kind intrinsic structure (for a more detailed discussion of this topic see Refs.~\cite{Krasnov2022,*Krasnov2022a}).

Therefore, the reasonable conditions, under which the vacuum positron emission can be unambiguously detected on the nuclear conversion pairs background, should play an exceptional role  in slow  ions collisions, aimed at the search of such events. However, the estimates for $Z^{\ast}$, achieved above, show that for such conditions two heavy ion collisions are out of work. The gap between the highest achievable nowadays $Z=192$ for Cm+Cm colliding beams and the estimate from below $Z^{\ast} \simeq 250-260$ is to large for any kind special efforts in searching for the signal of spontaneous emission in such collisions. Note also that with further growth of $Z$ the decline  of $ \E_{VP}^{ren}(Z,R)$ into the negative range with decreasing $R$  becomes more and more pronounced, as it is clearly seen in Fig.\ref{VP500(R)}, and so  the total rate of emission becomes significantly more high. It happens because the decay of the Dirac sea VP-energy $ \E_{D, VP}^{ren}(Z,R)$ at $R=R_{min}(Z)$ in the range $300 < Z < 600$ proceeds extremely rapidly ~\cite{Grashin2022a,*Krasnov2022a}. In particular, $ \E_{D, VP}^{ren}(Z,R_{min}(Z)$ for $Z=300$ is only $\simeq - 11$ MeV, while for $Z=600$ it exceeds $- 500$ MeV. At such a rate of VP-energy decay into the negative range the spontaneous emission has every chance of being detected against the background of a conversion pairs that grows $\sim Z^2$.

  One possible way to access QED at a supercritical Coulomb field is to consider more complicated collisions, which are based on the regular polyhedron symmetry --- synchronized slow heavy ion beams move  from the vertices of the polyhedron towards its center. The simplest collision of such kind contains 4 beams, arranged as the mean lines of the tetrahedron with the intersection angle $\vt=2 \arcsin (\sqrt{2/3}) \simeq 109^{\small 0}\,28'$, akin to s-p hybridized tetrahedron carbon bonds. The next and probably the most preferable one is the 6-beam configuration, reproducing  6 semi-axes of the rectangular coordinate frame. The advantage of such many-beam collisions is evident. On the other hand, such collisions require serious additional efforts for their implementation. Therefore, an additional tasty candy is needed to arouse sufficient interest in such a project, and it is really possible, but this issue requires for a special discussion.

\section{Acknowledgments}

The authors are very indebted to Dr. Yu.S.Voronina, Dr. O.V.Pavlovsky   and A.A.Krasnov from MSU Department of Physics and to Dr. A.S.Davydov from Kurchatov Institute and Dr. A.A. Roenko from JINR (Dubna) for interest and helpful discussions.  This work has been supported in part by the RF Ministry of Sc. $\&$ Ed.  Scientific Research Program, projects No. 01-2014-63889, A16-116021760047-5, and by RFBR grant No. 14-02-01261.  The research is carried out using the equipment of the shared research facilities of HPC computing resources at Moscow Lomonosov State University, project No. 2226. A large amount (more than 50 $\%$) of numerical calculations has been performed using computing resources of the federal
collective usage center Complex for Simulation and Data Processing for
Mega science Facilities at NRC Kurchatov Institute, research group No. g-0142.

\begin{widetext}

\section{Appendix A. Explicit formulae for the Born term  (\ref{3.261})
in the case of the potential (\ref{1.5a})}

Here we list the results of integration over $dr'$ in the Born term (\ref{3.261})  for  the potential (\ref{1.5a}) in the s-channel ($k=1$), which are required for the main part of the work. For these purposes we define first
\beq \begin{gathered}
\label{A.1}
a_1(y,r)={\sh\[2 \g r\] - 2 \g r  \over 2\,\pi\, \g^2 } \ , \quad
a_2(y,r)={2+2 (\g r)^2 -2 \ch\[2 \g r\] + \g r \sh\[ 2 \g r\] \over 2\,  \pi\, r \g^3 } \ , \\
c_1(y,r)={\mathrm{Chi}\[2\, \g\, r\]-\mathrm{Chi} \[2\, \g\, R\] + \ln\[R/r\] \over \pi \g } \ , \\
c_2(y,r)={1 \over 2\, \pi \g^3}\, \Big[1/r^2-1/R^2+\ch\[2 \g R \]/R^2-\ch\[2 \g r\]/r^2 \ + \\ + \ 2\, \g^2 \(\mathrm{Chi}\[2\, \g R\]-\mathrm{Chi}\[2\, \g r\] - \ln\[\g R\]+ \ln\[\g r\]\) +  2\,  \g \(\sh\[2 \g r\]/r-\sh\[2 \g R\]/R\) \Big] \ , \\
A_1(y,r)=\tt(R-r)\,a_1(y,r)/R +  \(a_1(y,R)/R+c_1(y,r)\)\,\tt(r-R) \ , \\
A_2(y,r)=\tt(R-r)\,a_2(y,r)/R +  \(a_2(y,R)/R+c_2(y,r)\)\,\tt(r-R) \ , \\
b_1(y,r)=\pi\,\[\mathrm{e}^{-2 \g r }-\mathrm{e}^{-2 \g R}\]/4 \g^2 \ , \quad
b_2(y,r)={R\, \mathrm{e}^{2 \g R } (2+ \g r )-\pi\,r\, \mathrm{e}^{-2 \g R}  (2+\g R) \over 4\, R  r \g^3} \ ,\\
d_1(y,r)=\pi\, \G \[0,2\, \g r\] /2 \g \ , \quad
d_2(y,r)=\pi\, {\mathrm{e}^{-2 \g r } (1+2 \g r)/2 r^2 + \g^2 \mathrm{Ei}\[-2 \g r \] \over 2 \g^3} \ , \\
B_1(y,r)=\tt(R-r)\,\(b_1(y,r)/R+d_1(y,R)\) +  d_1(y,r)\,\tt(r-R) \ , \\
B_2(y,r)=\tt(R-r)\,\(b_2(y,r)/R+d_2(y,R)\) +  d_2(y,r)\,\tt(r-R) \ .
 \end{gathered}\eeq
By means of this set of subsidiary functions the final expression for the Born term in the s-channel reads
\begin{multline}
\label{A.2}
 \tr G^{(1)}_{1}(r;i y) =  {Q \over r}\, \Bigg[(1-y^2)\,\(K^2_{1/2}(\g r)\, A_1(y,r) \ + \ K^2_{3/2}(\g r)\, A_2(y,r) +  I^2_{1/2}(\g r)\,B_1(y,r) \ + \ I^2_{3/2}(\g r)\,B_2(y,r) \)
+   \\  +
 (1+y^2)\,\(K^2_{1/2}(\g r)\, A_2(y,r) \ + \ K^2_{3/2}(\g r)\, A_1(y,r) +  I^2_{1/2}(\g r)\,B_2(y,r) \ + \ I^2_{3/2}(\g r)\,B_1(y,r) \)\Bigg] \ .
\end{multline}

Quite analogously, the integration over $dr'$ in the Born term (\ref{3.261}) can be performed for any $k>1$, leading to the corresponding expression for the Born term in a compact form, but with growing $k$ this calculation becomes more and more cumbersome. However, within the present analysis we use only the s-channel, which is defined via eqs.(\ref{A.1}-\ref{A.2})).
And although these expressions contain integral exponent and cosine functions, they are simple enough to provide a quite effective numerical integration over $dy$, required to extract the most important nonlinear component  $\vr^{(3+)}_{VP,|k|}(r)$ from  the total partial VP-density.
\end{widetext}

\section{Appendix B. The behavior of the Born term  (\ref{3.261}) in various asymptotical regimes}

Here we explore in the general form the asymptotics of the Born term, which plays an important role in the content of Sect. III and IV. For these purposes we recall first the asymptotics of $I_\n(z)$ and $K_\n(z)$
\beq\label{B.1a}
 z \to \inf \ , \quad I_\n(z) \to {\mathrm{e}^z \over \sqrt{2 \pi z}}  \ , \quad K_\n(z) \to \sqrt{{\pi \over 2 z}}\, \mathrm{e}^{-z} \ ,
\eeq
\beq\begin{gathered}\label{B.1b}
z \to 0 \ , \quad I_\n(z) \to {(z/2)^\n \over \G(\n+1) }  \ ,  \\ K_{n+1/2}(z) \to \sqrt{{\pi \over 2 z}}\,{\G(2n+1) \over n!\, (2z)^n} \ .
\end{gathered}\eeq
For the asymptotics of the Born term for $r \to \inf$ it suffers to consider in details only the behavior of terms
\beq\label{B.2}
K^2_\n(\g r)\,\int\limits_0^r \! dr'\,r'V(r')\,I^2_\n(\g r')
\eeq and
\beq\label{B.3}
I^2_\n(\g r)\,\int\limits_r^{\inf} \! dr'\,r'V(r')\,K^2_\n(\g r')
 \eeq
 for large $r$, since this asymptotics does not depend on $\n$. In this limit, we can freely replace  $r'V(r') \to -Q$, and thereafter, upon inserting the asymptotics (\ref{B.1a}) and integrating the inner integral over $dr'$  by parts in such a way that the exponent $\mathrm{e}^{2 \g r'}$ is integrated, while the remaining inverse powers of $r'$ are differentiated, one obtains from  (\ref{B.2}) the leading term in the  form $(-Q/ (2\g)^3\,r^2)$. The same result follows from  (\ref{B.3}). Collecting all the leading terms of such type in (\ref{3.261}) together, one finds finally
\beq\label{B.4}
\tr G^{(1)}_{k}(r;i y) \to Q/ (\g\,r)^3 \ , \quad r \to \inf \ .
\eeq
Moreover, this asymptotics is uniform in $y$, since $\g \geq 1$ for the whole range $0\leq y \leq \inf$ .

The asymptotics of the Born term for $r \to 0$ is more subtle, since different terms in the total expression (\ref{3.261}) reveal different behavior in this limit. When $y$ is finite, then the main contribution is produced by the following terms in  square brackets in the r.h.s. of (\ref{3.261})
\begin{multline}
\label{B.5}
(1+y^2)\,\(K^2_{k+1/2}(\g r)\,\int\limits_0^r \! dr'\,r'V(r')\,I^2_{k-1/2}(\g r') \ +  \right. \\ \left.
 + \ I^2_{k-1/2}(\g r)\,\int\limits_r^{\inf} \! dr'\,r'V(r')\,K^2_{k+1/2}(\g r') \) \ ,
\end{multline}
where $V(r')$ should be now replaced by the limiting value of the external Coulomb potential under question $V(0)$. Upon integrating the inner integrals over $dr'$ in (\ref{B.5}) with account of the asymptotics (\ref{B.1b})  one finds for the limit of the expression (\ref{B.5})
\beq\label{B.6}
{\pi\, V(0) \over 2^{4k}}\,{4 k \over 4 k^2 -1}\,{\G^2(2k+1) \over \G^2(k+1)\,\G^2(k+1/2)} \ .
\eeq
It is easy to verify that remaining  two terms in the r.h.s. of (\ref{3.261}) with the same multiplier $(1+y^2)$ reveal the asymptotics $O(r^2)$, while the terms with multiplier $(1-y^2)$ turn out to be $O(r)$, correspondingly. So in this case  the asymptotics of the Born term for $r \to 0$ turns out to be
\beq\label{B.7}
\tr G^{(1)}_{k}(r;i y) \to - C_k\,{ V(0)\over r} \ ,
\eeq
with $C_k$ being a constant depending only on the angular number $k$.

In the regime when $r \to 0$, but simultaneously $ y \to \inf$ in such a way that $t=\g r$ remains finite and non-zero, one should first perform the  replacements  $V(r') \to V(0)\, , 1 \pm y^2 \to \pm \g^2$, and thereafter transform the expression (\ref{3.261}) into
 \begin{widetext}
\beq \begin{gathered}
\label{B.8}
 \tr G^{(1)}_{k}(r;i y) \to - {V(0) \over r}\, \Bigg[\(K^2_{k-1/2}(t)-K^2_{k+1/2}(t)\)\,\int\limits_0^t \! dt'\,t'\,\(I^2_{k-1/2}(t')-I^2_{k+1/2}(t')\) +  \\
 + \(I^2_{k+1/2}(t)-I^2_{k-1/2}(t)\)\,\int\limits_t^{\inf} \! dt'\,t'\,\(K^2_{k-1/2}(t')-K^2_{k+1/2}(t')\)\Bigg] \ ,
\end{gathered}\eeq
 \end{widetext}
whence there follows again the asymptotics of $\tr G^{(1)}_{k}(r;i y)$  in the form (\ref{B.7}). However, when $y$ grows further and so  $t \to \inf$, the r.h.s. of (\ref{B.8}) vanishes. In this case the main contribution to the asymptotics of $\tr G^{(1)}_{k}(r;i y)$ comes from units in expressions $1 \pm y^2$, which have been  dropped in the previous calculation, and reads
\beq\label{B.9}
\tr G^{(1)}_{k}(r;i y) \to - V(0) r/t^3 \ .
\eeq

The behavior of the Born term for $y \to \inf$ with $r$ being finite and non-zero follows from the initial expression  (\ref{3.261}), where  it again suffers to consider in details only the  terms (\ref{B.2}) and (\ref{B.3}) for large $\g r$. In this limit we can freely replace  $K^2_\n(\g r)\,, K^2_\n(\g r')\,,I^2_\n(\g r)\,, I^2_\n(\g r')$ by the asymptotics (\ref{B.1a}), but $r'V(r')$ in the inner integral must now be kept unchanged. Thereafter, upon integrating the inner integrals over $dr'$  by parts one obtains from  (\ref{B.2}) the leading term in the  form $V(r)/ (2\g)^3\,r$. The same result follows from  (\ref{B.3}). Collecting all the leading terms of such type in (\ref{3.261}) together, one finds finally
\beq\label{B.10}
\tr G^{(1)}_{k}(r;i y) \to -V(r)/ (\g^3\,r^2) \ , \quad y \to \inf \ , \quad 0< r<\inf .
\eeq
For $r \to \inf$, upon replacing $V(r)$ by the general asymptotics $V(r) \to -Q/r$,  one obtains from (\ref{B.10}) the former result (\ref{B.4}), while for $r \to 0\, , V(r) \to V(0)$  in this limit there appears   firstly  the regime $t=\g r \to \inf$, where (\ref{B.10}) transforms into (\ref{B.9}). When $r$ continues to decrease further so that $t=\g r $ becomes finite or even tends to zero, then the treatment should be reversed to the case $r \to 0$  considered above.

\begin{figure*}[ht!]
\subfigure[]{
		\includegraphics[width=0.8\columnwidth]{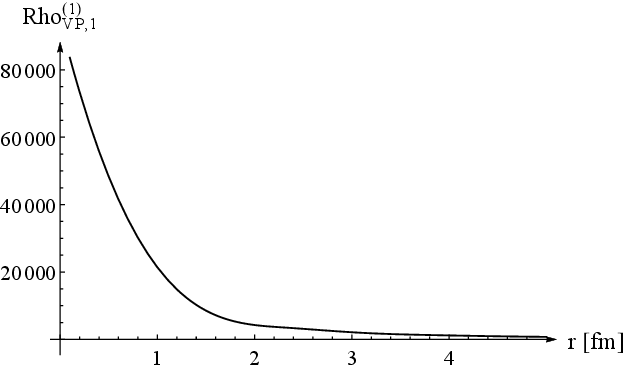}
}
\hfill
\subfigure[]{
		\includegraphics[width=0.8\columnwidth]{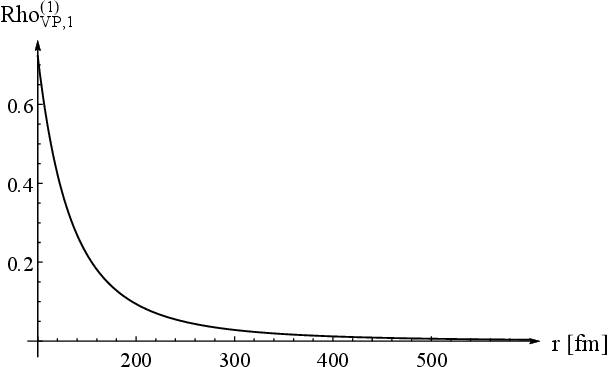}
}
\vfill
\subfigure[]{
		\includegraphics[width=1\columnwidth]{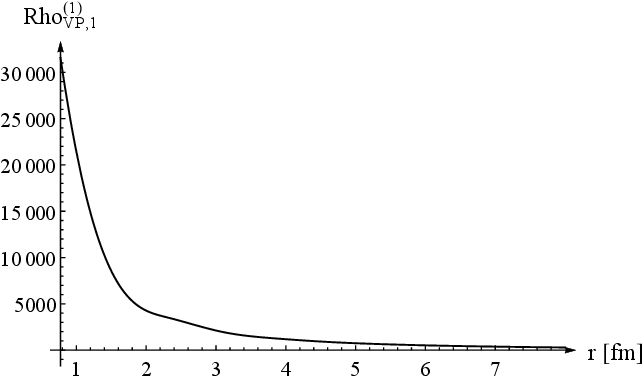}
}
\vfill
\subfigure[]{
		\includegraphics[width=0.8\columnwidth]{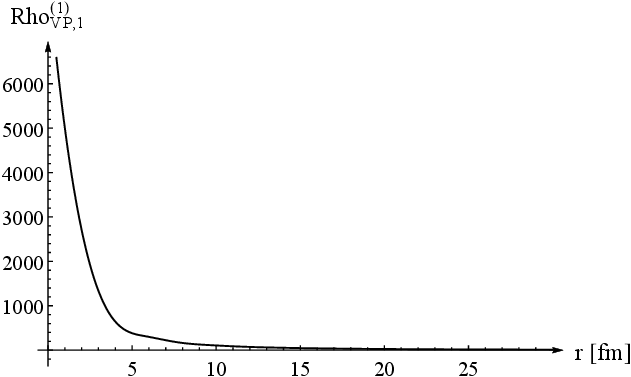}
}
\hfill
\subfigure[]{
		\includegraphics[width=0.8\columnwidth]{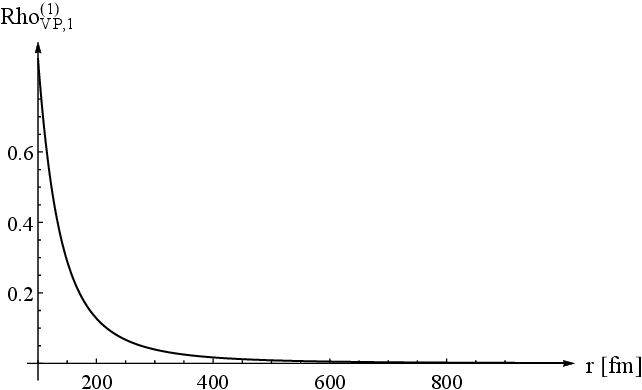}
}
\vfill
\subfigure[]{
		\includegraphics[width=1\columnwidth]{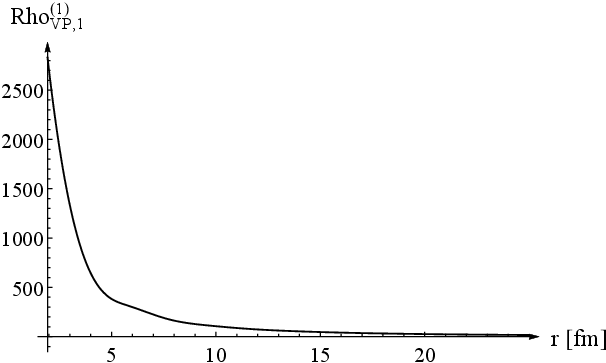}
}
\caption{The behavior of  $ \vr_{VP}^{(1)}(Z,r)$ in the $s$-channel, which for such $Z$  is the dominating one and contains more than $99 \%$ of contributions from all the VP-effects under question  for: (a-c) $Z=200$ with $R=0.1$; (d-f) $Z=280$ with $R=0.25$ .  }
	\label{Rho1_Z=200-280(r)}	
\end{figure*}

\section{Appendix C. The behavior of the first-order charge density $ \vr_{VP}^{(1)}(Z,r)$}

For our purposes it suffers to explore the behavior of the first-order charge density $ \vr_{VP}^{(1)}(Z,r)$  for the potential  (\ref{1.5a}) in the s-channel. The calculation is based on the explicit  formulae (\ref{A.1},\ref{A.2}), received in App.A. Although these expressions contain integral exponent and cosine functions and so cannot be integrated over $dy$ analytically, they are simple enough to provide a quite effective numerical integration. Such integration is performed via interpolation of the integrand by means of the grid containing 800-1000 segments on the interval $0 < y < 25000$, which is quite sufficient for the considered range $100 < Z < 300$. The restoration  of $\vr^{(1)}_{VP,1}(Z,r)$ as a radial function is based on the grid, which  contains about 100 points on the interval $0< r < 15$, with subsequent smooth interpolation. For the range $100 < Z < 300$ such interval is quite sufficient, since the density decreases quite rapidly, while the number of radial points in the grid is enough for precise restoration of $\vr^{(1)}_{VP,1}(Z,r)$ for such $Z$. The results of calculation are shown in Figs.\ref{Rho1_Z=200-280(r)} for two quite representative values of $Z=200\,  , 280$. The corresponding radii of the Coulomb source are chosen quite large to make clear the details of nonlinear behavior of density in the intermediate range of $r$, where the nonlinear transformation of $O(1/r)$ asymptotics for $r \to 0$ into $O(1/r^3)$ asymptotics for $r \to \inf$ takes place. For each $Z$ the density is given in three  different areas to show the  density distribution at small, intermediate and large values of the radial variable. Remarkably enough, the general features of curves for different $Z$ are quite similar. They are strictly positive and differ mostly in magnitudes at small and intermediate values of the radial variable.

\begin{figure*}[ht!]
\subfigure[]{
		\includegraphics[width=1.9\columnwidth]{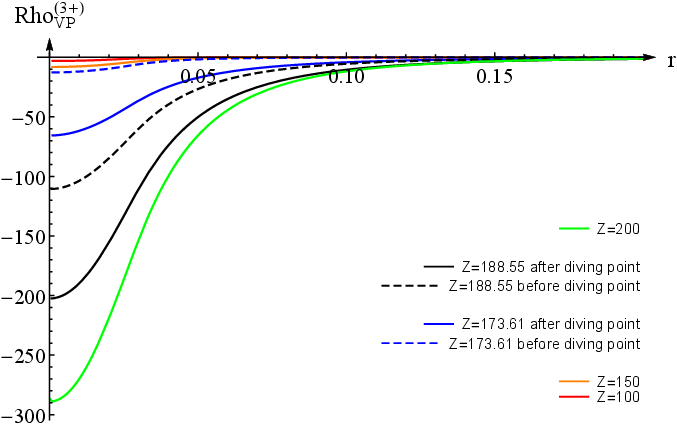}
}
\vfill
\subfigure[]{
		\includegraphics[width=1.9\columnwidth]{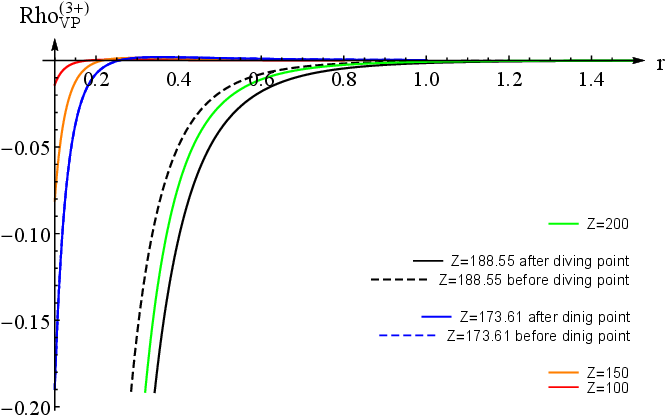}
}
\caption{(Color online) The behavior of  $ \vr_{VP}^{(3+)}(Z,r)$ in the $s$-channel, which for such $Z$  is the dominating one and contains more than $99 \%$ of contributions from all the VP-effects under question  for: (a) the range $0<r<0.2$, where the VP-density varies most pronouncedly; (b) the range $0.1<r<1.5$ to show the asymptotic tail of the density distribution.  }
	\label{Rho3+Z=100-200(r)}	
\end{figure*}

\begin{figure*}[bt!]
\subfigure[]{
		\includegraphics[width=1.8\columnwidth]{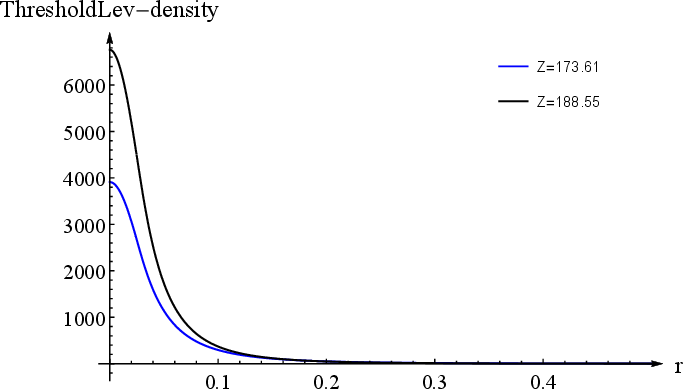}
}
\caption{(Color online) The normalized to unity radial profiles $|\p_{\n} (r)|^2$  of dived levels (per spin projection) at the threshold of the lower continuum    for $\n=1s\,,2p$. The colors of $|\p_{1s} (r)|^2$ and $|\p_{2p} (r)|^2$ are intentionally chosen in correspondence with VP-densities of diving levels $1s$ (blue) and $2p$ (black). }
\label{threholdlev-density1s2p(r)}	
\end{figure*}

\begin{figure*}[ht!]
\subfigure[]{
		\includegraphics[width=1.8\columnwidth]{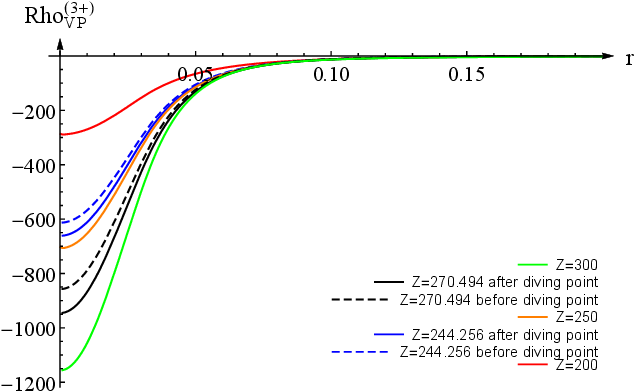}
}
\vfill
\subfigure[]{
		\includegraphics[width=1.8\columnwidth]{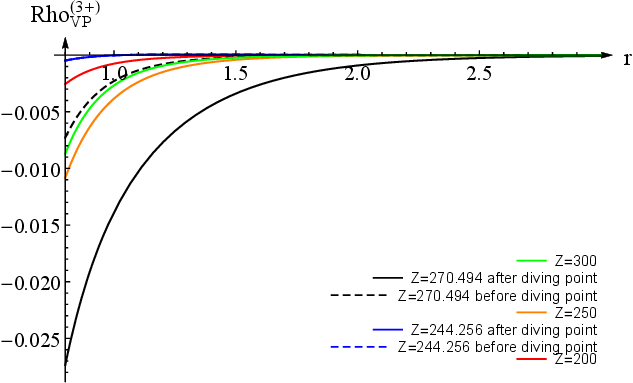}
}
\caption{(Color online) The behavior of  $ \vr_{VP}^{3+}(Z,r)$ in the $s$-channel, which for such $Z$  is the dominating one and contains more than $99 \%$ of contributions from all the VP-effects under question  for:(a) the range $0<r<0.2$, where the VP-density varies most pronouncedly ; (b) the range $0.8<r<3.0$ to show the asymptotic tail of the density distribution.  }
	\label{Rho3+Z=200-300(r)}	
\end{figure*}

\begin{figure*}[bt!]
\subfigure[]{
		\includegraphics[width=1.8\columnwidth]{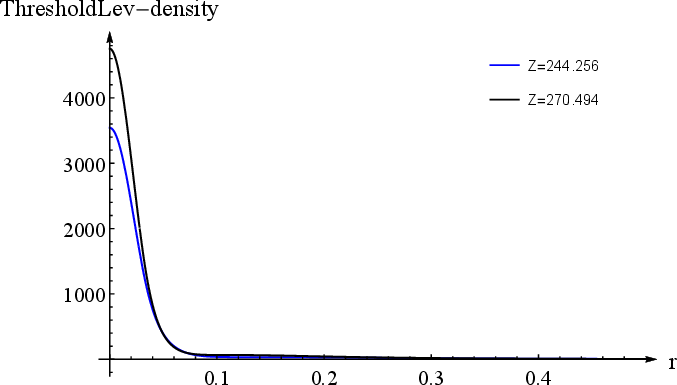}
}
\caption{(Color online) The normalized to unity radial profiles $|\p_{\n} (r)|^2$  of dived levels (per spin projection) at the threshold of the lower continuum    for $\n=2s\,,3p$. The colors of $|\p_{2s} (r)|^2$ and $|\p_{3p} (r)|^2$ are intentionally chosen in correspondence with VP-densities of diving levels $2s$ (blue) and $3p$ (black). }
\label{threholdlev-density2s3p(r)}	
\end{figure*}

\begin{figure*}[ht!]
\subfigure[]{
		\includegraphics[width=1.8\columnwidth]{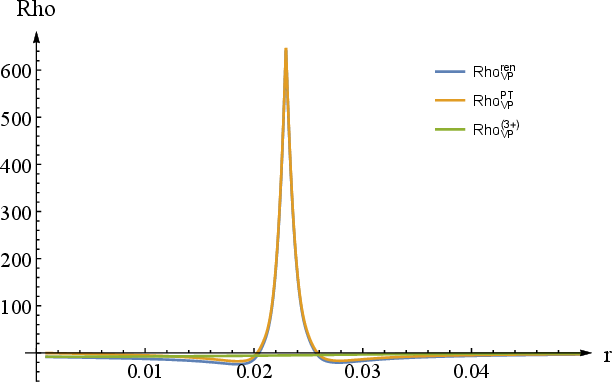}
}
\vfill
\subfigure[]{
		\includegraphics[width=1.8\columnwidth]{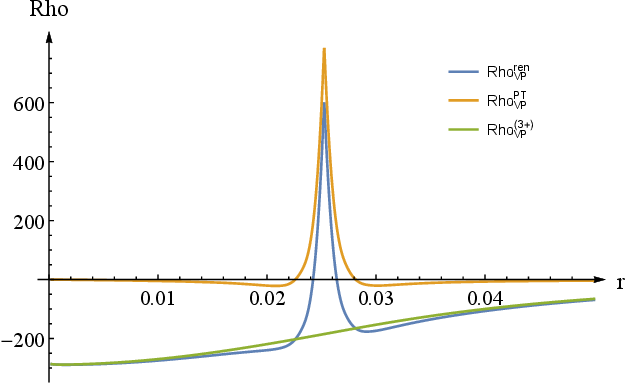}
}
\caption{(Color online) The behavior of  $ \vr_{VP}^{ren}(Z,r)$ in the $s$-channel, which for such $Z$  is the dominating one and contains more than $99 \%$ of contributions from all the VP-effects under question  for: (a) $Z=150$, when there are no dived levels yet; (b) $Z=200$, when the first two levels $1s$- and $2p$- have already dived into the lower continuum. }
	\label{Rho_ren_Z=150-200(r)}	
\end{figure*}
\begin{figure*}[bt!]
\subfigure[]{
		\includegraphics[width=1.8\columnwidth]{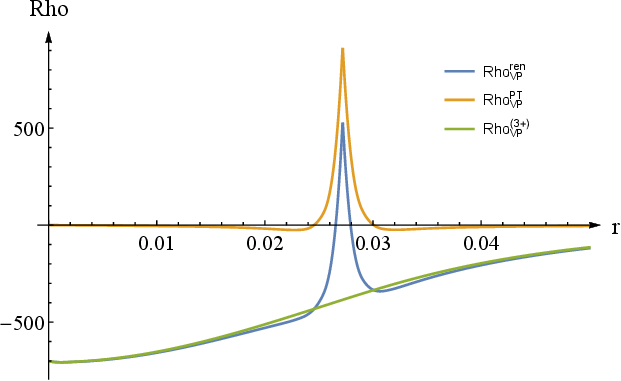}
}
\vfill
\subfigure[]{
		\includegraphics[width=1.8\columnwidth]{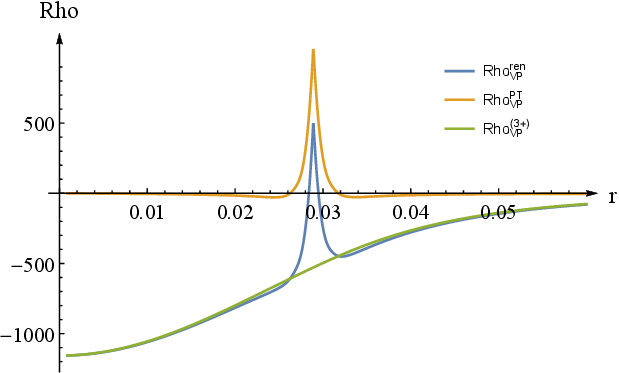}
}
\caption{(Color online) The behavior of  $ \vr_{VP}^{ren}(Z,r)$ in the $s$-channel, which for such $Z$  is the dominating one and contains more than $99 \%$ of contributions from all the VP-effects under question  for: (a) $Z=250$, when three levels $1s\,,2p\,,2s$ have already dived into the lower continuum; (b) $Z=300$ with four dived levels $1s\,,2p\,,2s\,,3p$. }
\label{Rho_ren_Z=250-300(r)}	
\end{figure*}

\section{Appendix D. The behavior of VP-charge density}

In the present context the behavior of VP-charge density is shown in Figs.\ref{Rho3+Z=100-200(r)}--\ref{Rho_ren_Z=250-300(r)} for the potential  (\ref{1.5a}) with the radius $R_{min}(Z)$ defined in (\ref{6.1}). The case of potential  (\ref{1.5a}) is more preferable for presentation since the corresponding partial Green functions can be  explicitly written in terms of Bessel and confluent hypergeometric functions, which gradually simplifies  the calculation of the integral over the imaginary axis in (\ref{3.21}). Explicit formulae for  $\tr G_{k}(r,r';i y)$  are given in Ref.~\cite{Gyulassy1975}.

 In this case in the range $0<Z<300$  there are four dived levels in the $s$-channel: $1s$ at $Z=173.61$, $2p$ at $Z=188.55$, $2s$ at $Z=244.256$ and $3p$ at $Z=270.494$. These values of critical charges are obtained from eq. (\ref{5.9}). In Figs.\ref{Rho3+Z=100-200(r)},\ref{Rho3+Z=200-300(r)} the behavior of the non-perturbative component $ \vr_{VP}^{(3+)}(Z,r)$ is presented. The jumps in VP-density,  accompanying each level diving, are clearly seen, while the direct calculation confirms that the corresponding jumps in the total VP-charge are equal indeed to  $(-2|e||k|)$. Moreover, the spatial  profiles of these jumps in VP-density coincide exactly with those of the dived levels  at the threshold of the lower continuum, multiplied by the same factor  $(-2|e||k|)$. The normalized to unity radial profiles $|\p_{\n} (r)|^2$  of dived levels (per spin projection) at the threshold of the lower continuum  are shown in Figs. \ref{threholdlev-density1s2p(r)}\,,\ref{threholdlev-density2s3p(r)}.

 It should be noted that the calculation of $ \vr_{VP}^{(3+)}(Z,r)$ with growing  $Z$ and $R$ turns out to be time consuming,  even for the potential  (\ref{1.5a}), and requires for a solid parallelizing.  Principally, the integration over $dy$ is performed along the lines of App.C via interpolation of the integrand by means of the grid containing 800-1000 segments on the interval $0 < y < 25000$. Further restoration  of $\vr^{(1)}_{VP,1}(Z,r)$ as a radial function is performed on a radial grid, which  contains about 100 points on the interval $0< r < 15$, with subsequent smooth interpolation. For the range $100 < Z < 300$ such number of radial points in the grid is enough for a reliable restoration of $\vr^{(3+)}_{VP,1}(Z,r)$ for such $Z$ and $R$. However, for large enough $Z\sim 200-300$ and $R \sim 0.1 - 0.5$ the calculation of $\tr G^{(3+)}_{1}(r;i y)$ with reliable accuracy requires for a very high precision (about several thousands of  true characters) in  the values of confluent  hypergeometric functions with the main purely imaginary argument $2\, i \g R$ exceeding $ \sim i\, (10000-25000)$, which is a nontrivial task even for the modest soft- and hard-ware.

 In Figs.\ref{Rho_ren_Z=150-200(r)},\ref{Rho_ren_Z=250-300(r)} the behavior of the total renormalized VP-density component $ \vr^{ren}_{VP}(Z,r)$, defined in  (\ref{3.27}),  is shown for four quite representative values of $Z$, namely $Z=150\,,200\,,250\,,300$. To create the best performance of   presentation we use the following trick. For the potential (\ref{1.5a}) the perturbative renormalized VP-density $\vr^{PT}_{VP}(\vec{r})$ reveals a $\d$-like singularity at $r=R$, what is difficult to reproduce in a well visualized  picture. Therefore in these plots we replace the true perturbative VP-density by another one, created by a smoothed  $\d$-like external charge configuration of the form
\beq\label{D.1}
\vr_{ext}(r)= {Z\, |e| \over N(\l,R)}\,\( \mathrm{e}^{(r-R)/\l} \tt(R-r) + \mathrm{e}^{(R-r)/\l}\tt(r-R) \) \ ,
\eeq
where the normalization factor $N(\l,R)$ equals to
\beq\label{D.2}
N(\l,R)=8 \pi \l \[R^2+(2- \mathrm{e}^{-R/\l}) \l^2\] \ ,
\eeq
while the smoothing factor $\l$ is chosen in such a way, which provides the best performance of   the whole picture for VP-densities. In  Figs.\ref{Rho_ren_Z=150-200(r)},\ref{Rho_ren_Z=250-300(r)} the value of  $\l$ is chosen $1/1000$.

The  corresponding perturbative renormalized VP-density is calculated via
 \beq\label{D.3}
\vr^{PT}_{VP}(r)= Z |e|\,{ \a \over 2\, \pi^2\,r}\,\int_0^\inf \! q\,dq\,\sin(q r)\, S(q)\, \vr_{ext}(q) \ ,
\eeq
where
 \beq\label{D.4}
\vr_{ext}(q)= {\mathrm{e}^{R/\l} \(2 q \l^2 \cos(q R)+R (1+q^2 \l^2) \sin[q R]\) -q \l^2 \over q \(1+q^2 \l^2\)^2 \(\mathrm{e}^{R/\l} (R^2+2 \l^2)-\l^2\) } \ .
\eeq

At the same time, the non-perturbative component of VP-density $\vr^{(3+)}_{VP}(r)$ is calculated for the original potential (\ref{1.5a}). For sufficiently small $\l$ such combination of smoothed $\vr^{PT}_{VP}(r)$ and original $\vr^{(3+)}_{VP}(r)$ reproduces the true one quite correctly, but simultaneously is more appropriate for the visual study than the $\d$-like jump at $r=R$.  In particular, it is clearly seen, how the total VP-density transmutes from the one close to $\vr^{PT}_{VP}(r)$ at $Z=150$ with no dived levels and vanishing total VP-charge into the configurations with more and more dived levels and so with increasing role of  $\vr^{(3+)}_{VP}(r)$. Note also, that the asymptotics of the renormalized $\vr^{PT}_{VP}(r)$ for $r \to \inf$ tends to zero always from below, while that of $\vr^{(3+)}_{VP}(r)$ from above.

\bibliography{VP3DC}

\end{document}